\documentclass[twocolumn,epsf,prb,amsmath,amssymb,showpacs,superscriptaddress]{revtex4}
\usepackage{graphicx}
\usepackage{amsmath}
\usepackage{amssymb}
\usepackage[T1]{fontenc}

\begin{document}
\title{Energy levels of triangular and hexagonal graphene quantum dots: a comparative study between the tight-binding and the Dirac equation approach}

\author{M. Zarenia}
\affiliation{Departement Fysica, Universiteit Antwerpen, \\
Groenenborgerlaan 171, B-2020 Antwerpen, Belgium}
\author{A. Chaves}
\affiliation{Departement Fysica, Universiteit Antwerpen, \\
Groenenborgerlaan 171, B-2020 Antwerpen, Belgium}
\affiliation{Departamento de F\'{\i}sica, Universidade
Federal do Cear\'a, Fortaleza, Cear\'a, $60455$-$760$, Brazil}
\author{G. A. Farias}
\affiliation{Departamento de F\'{\i}sica, Universidade
Federal do Cear\'a, Fortaleza, Cear\'a, $60455$-$760$, Brazil}
\author{F.~M.~Peeters}\email{francois.peeters@ua.ac.be}
\affiliation{Departement Fysica, Universiteit Antwerpen, \\
Groenenborgerlaan 171, B-2020 Antwerpen, Belgium}
\affiliation{Departamento de F\'{\i}sica, Universidade
Federal do Cear\'a, Fortaleza, Cear\'a, $60455$-$760$, Brazil}

\begin{abstract}
The Dirac equation is solved for triangular and hexagonal graphene
quantum dots for different boundary conditions in the presence of a
perpendicular magnetic field. We analyze the influence of the dot
size and its geometry on their energy spectrum. A comparison between
the results obtained for graphene dots with zigzag and armchair
edges, as well as for infinite-mass boundary condition, is presented
and our results show that the type of graphene dot edge and the
choice of the appropriate boundary conditions have a very important
influence on the energy spectrum. The single particle energy levels
are calculated as function of an external perpendicular magnetic
field which lifts degeneracies. Comparing the energy spectra
obtained from the tight-binding approximation to those obtained from
the continuum Dirac equation approach, we verify that the behavior
of the energies as function of the dot size or the applied magnetic
field are qualitatively similar, but in some cases quantitative
differences can exist.
\end{abstract}
 \maketitle

\section{Introduction}

Since its recent discovery \cite{novo1}, graphene (a single layer of
carbon atoms) has been attracting a lot of interest, due to its
unique band structure, which is gapless and exhibits an
approximately linear dispersion relation at two inequivalent points
of the reciprocal space (labeled as $K$ and $K'$) in the vicinity of
the Fermi energy. The linearity of the band structure allows one to
describe the carriers close to the $K$ and $K'$ points in a
continuum model, using the Dirac equation with massless
particles\cite{CastroNeto}. Because of the well known Klein
tunneling effect in graphene, which prevents electrical confinement
of electrons, the lateral confinement of Dirac carriers is a big
challenge in manufacturing graphene-based electronic
devices\cite{Katsnelson, Milton, Matulis}. Different suggestions
have been made to realize lateral confinement of electrons in
graphene, e.g. by means of gap engineering, provided by a space
dependent mass term, \cite{GiavarasNori1, GiavarasNori2} or,
alternatively, by combining an external magnetic field
\cite{Giavaras} or a finite mass term \cite{Recher} with an
electrostatic potential. On the other hand, recent improvements of
different fabrication techniques made possible cutting and
manufacturing of single layer graphene flakes, with different shapes
and sizes\cite{Novoselov3, Hiura, Subramanian}, where such a lateral
confinement naturally occurs. Using the tight-binding model (TBM),
remarkable effects have been reported as a consequence of the type
of the edges and the geometry of these
flakes\cite{Ezawa,Guclu,Heiskanen,Akola,Zhang,Kosimov,Ligia2}: {\it
i}) zero-energy states are predicted for triangular graphene flakes
with zigzag boundaries, {\it ii}) for very small flakes a gap opens
(the energy gap of different graphene flakes was recently
investigated experimentally\cite{ritter}) and the density of states
(DOS) strongly depends on the type of the edges for any dot
geometry, and {\it iii}) the energy levels of graphene quantum dots
in the presence of a magnetic field approach the Landau levels with
increasing magnetic field.

\begin{figure}
\centering
\includegraphics[width=7.5cm]{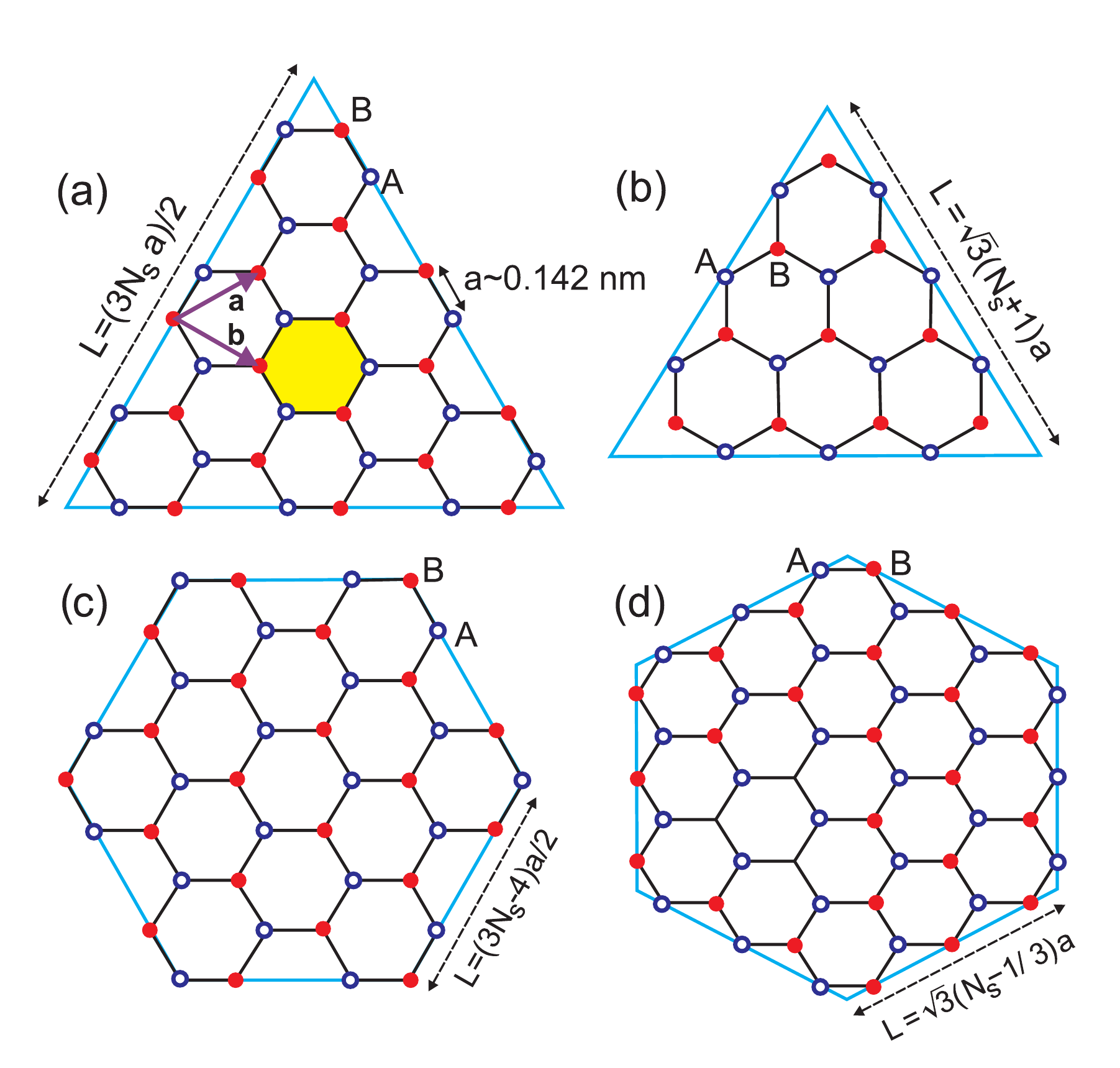}
\caption{(Color online) The lattice structure of triangular (upper
panels) and hexagonal (lower panels) graphene quantum dots with
(a,c) armchair edges and (b,d) zigzag edges. $a=0.142$ nm is the C-C
distance and the primitive lattice vectors are denoted by {\bf a}
and {\bf b}. The atoms of the two sublattices are represented by
blue circles and red dots. The yellow region indicates the area of
one carbon hexagon. $N_{s}$ is the number of C-atoms in each side of
the dot.} \label{fig0}
\end{figure}

Recently, analytical results were reported for infinite mass
boundary conditions for circular disks, \cite{Schnez} for
triangular flakes with armchair edges\cite{Rozhkov} and zigzag
edges\cite{Guclu} and for square graphene quantum dots\cite{Tang}.
However, it is not always clear how the complicated boundary
conditions describing the zigzag and armchair edges can be invoked
in the continuum model. Furthermore, the geometry of the triangular
and hexagonal graphene flakes, make such systems harder to be
studied by analytical means. One has to rely on, either a
tight-binding model or a numerical solution of coupled differential
equations in case of the continuum model.

The continuum model describes very well the low energy states in an
infinite graphene sheet, but it is not clear if this is still the
case for small graphene flakes. Therefore, it is important to learn
if there is a minimum size beyond which the continuum model no
longer gives reliable predictions. Furthermore, because of the large
influence of the type of edges on the energy spectrum, and since it
is not always clear which boundary conditions should be invoked in
the Dirac equation for each possible geometry of the flake, a
comparison between the results obtained with the different possible
boundary conditions and a link with the TBM is an interesting issue
which requires a detailed study.

In this paper, by solving the Dirac equation numerically, we
present a theoretical study of the energy spectra of triangular
and hexagonal graphene quantum dots, where three types of boundary
conditions are invoked, namely, zigzag, armchair and infinite mass boundary
condition. The influence of an external magnetic
field, perpendicular to the graphene layer, on the energy spectrum
of the quantum dots is also analyzed. A comparison between the
results obtained with the continuum model and those obtained from
the tight-binding approach will be made.

This paper is organized as follows. In Sec. II we present
a brief outline of the tight-binding model (TBM). The model based on
the Dirac-Weyl equation is presented in Sec. III and the different boundary
conditions are separately analyzed in this section. Our numerical
results are reported in Sec. IV. The summary and
conclusions of this work are presented in Sec. V.

\section{Tight-binding model}
The tight-binding Hamiltonian within the nearest neighbor approximation is
\begin{equation}\label{eq_ham_TB}
H=\sum_{n}E_{n}c_{n}c_{n}^{\dag}+\sum_{<n,m>}(t_{n,m}c_{n}^{\dag}c_{m} + h. c.),
\end{equation}
where $E_{n}$ is the energy of the $n$-th site, $t_{n,m}$ is the
hopping energy and $c_{n}^{\dag}$ ($c_{n}$) is the creation
(annihilation) operator of the $\pi$ electron at site $n$.
Note that, for each site $n$, the summation is taken over all
nearest neighboring sites $m$. In the presence of a magnetic
field, the transfer energy becomes $\displaystyle{t\rightarrow
te^{i2\pi\Phi_{n,m}}}$, where
$\Phi_{n,m}=(1/\Phi_{0})\int_{r_n}^{r_m}\mathbf{A}\cdot
\mathbf{dl}$ is the Peierls phase, with $\Phi_{0}=h/e$ the
magnetic quantum flux and $\mathbf{A}$ the vector potential.

Triangular and hexagonal quantum dots with zigzag and armchair edges
are illustrated in Fig. 1, where the vectors
$\mathbf{a}=a(3/2,\sqrt{3}/2)$ and $\mathbf{b}=a(3/2,-\sqrt{3}/2)$,
with $a = 0.142$ nm the lattice parameter (or the C-C distance), are
introduced as primitive lattice vectors. In the present work, we
will consider only the interaction between each atom $n$ and its
three first nearest neighbors. In the case of graphene, this
interaction has the hopping energy $t=2.7$ eV. The vector potential
corresponding to the external magnetic field $\mathbf{B} = B\hat{z}$
perpendicular to the layer is chosen as the Landau gauge
$A=(0,Bx,0)$. With this choice of gauge, the Peierls phase for a
transition between two sites $n$ and $m$ is $\Phi_{n,m} = 0$ in the
$x$ direction and $\Phi_{n,m} = \pm(x/3a)\Phi_c/\Phi_0$ along the
$\pm y$ direction, where $\Phi_c=3\sqrt{3}a^{2}B/2$ is the magnetic
flux threading one carbon hexagon (the area of one carbon hexagon is
shown in Fig. \ref{fig0}(a) by the yellow region). An external
potential is represented by a variation in the on-site energies
$E_n$, and a vacancy or defect can be represented by setting the
energy of the vacant site to a larger value and the hopping terms to
these atoms as zero \cite{Ligia}. The Hamiltonian $H$ in Eq.
(\ref{eq_ham_TB}) can be represented in matrix form and the
eigenvalues and eigenfunctions of a graphene flake can be obtained
by diagonalization of the matrix.

Notice that the hexagonal lattice presented in Fig. 1 is not a
Bravais lattice, but a combination of two triangular lattices
composed by atoms labeled as type A (blue) and type B (red).
Accordingly, the tight-binding Hamiltonian of Eq. (\ref{eq_ham_TB})
can be rewritten as
\begin{equation}\label{eq_ham_TB2}
H=\sum_{n}E_n^Aa_{n}^{\dag}a_{n}+\sum_{n}E_n^Bb_{n}^{\dag}b_{n}+\sum_{<n,m>}(t_{n,m}a_{n}^{\dag}b_{m} + h.c.)
\end{equation}
where the operators $a_{n}^{\dag}$ ($a_{n}$) and $b_{n}^{\dag}$
($b_{n}$) create (annihilate) an electron in site $n$ of
lattice A and B, respectively.

\section{Continuum model: Dirac-Weyl equation}

Considering an infinite (periodic) graphene sheet and after, performing a
Fourier transform on the operators in Eq. (\ref{eq_ham_TB}) and
diagonalizing the resulting Hamiltonian leads to an energy
dispersion \cite{CastroNeto}
\begin{equation}\label{eq_fulldisp}
E(\mathbf{k})=\pm t \sqrt{3+2\cos(\sqrt{3}k_y
a)+4\cos\left(\frac{\sqrt{3}a}{2}k_y\right)\cos\left(\frac{3a}{2}k_x
\right)}.
\end{equation}
The first Brillouin zone in reciprocal space is a hexagon with
six Dirac points, where only two of them are inequivalent. From
the primitive vectors, we can find the position of these as
$\displaystyle{K=(2\pi/3a,2\pi/3\sqrt{3}a)}$ and
$\displaystyle{K'=(2\pi/3a,-2\pi/3\sqrt{3}a)}$. The states near
these points have approximately a linear dispersion and can be
described as massless Dirac fermions by the Hamiltonian
\begin{equation}\label{H1}
H=\left(
  \begin{array}{cc}
  H_{K} &\mathbf{0} \\
  \mathbf{0}&H_{K'} \\
  \end{array}
\right),
\end{equation}
where $H_{K}$ ($H_{K'}$) is the Hamiltonian in the $K$ ($K'$) point, which are given by
\begin{subequations}
\begin{equation}
H_{K}=v_{F}~\boldsymbol{\sigma}.~\mathbf{p},
\end{equation}
\begin{equation}
H_{K'}=v_{F}~\boldsymbol{\sigma^\ast}.~\mathbf{p},
\end{equation}
\end{subequations}
where $\boldsymbol{\sigma}=(\sigma_{x},\sigma_{y})$ are Pauli
matrices and $\boldsymbol{\sigma^\ast}=(\sigma_{x},-\sigma_{y})$ denotes the complex conjugate of the
matrix $\boldsymbol{\sigma}$. In the presence of a magnetic field
$B$ perpendicular to the graphene layer and using the Landau gauge,
one can simply rewrite Eq. (\ref{H1}) in the following form:
\begin{equation}\label{H3}
H=\left(
    \begin{array}{cccc}
      0 & \Pi_{-} & 0 & 0 \\
      \Pi_{+} & 0 & 0 & 0 \\
      0 & 0 & 0 & \Pi_{+} \\
      0 & 0 & \Pi_{-} & 0\\
    \end{array}
  \right),
\end{equation}
where,
\begin{equation}\label{eq}
\Pi_{\pm}=-i\hbar v_{F}\left[\frac{\partial}{\partial x}\pm i\frac{\partial}{\partial y}\mp\frac{2\pi B}{\Phi_0}x\right].
\end{equation}
The wave function in real space for the sublattice A is
\begin{subequations}
\begin{equation}\label{eqA}
\psi_{A}(\mathbf{r})= e^{i\mathbf{K}\cdot\mathbf{r}}\varphi_{A}(r) +
e^{i\mathbf{K'}\cdot\mathbf{r}}\varphi_{A'}(r),
\end{equation}
and for sublattice B it is given by,
\begin{equation}\label{eqB}
\psi_{B}(\mathbf{r})= e^{i\mathbf{K}\cdot\mathbf{r}}\varphi_{B}(r) + e^{i\mathbf{K'}\cdot\mathbf{r}}\varphi_{B'}(r).
\end{equation}
\end{subequations}
The Hamiltonian of Eq. (\ref{H3}) acts on the four-component wave
function
$\Psi=[\varphi_{A},\varphi_{B},\varphi_{A'},\varphi_{B'}]^{T}$,
which leads to the four coupled first-order differential equations:
\begin{subequations}\label{eqS}
\begin{eqnarray}
&&-i\left[\frac{\partial}{\partial x'}-i\frac{\partial}{\partial
y'}+\beta x'\right]\varphi_{B}=\epsilon \varphi_{A}, \\
&&-i\left[\frac{\partial}{\partial x'}+i\frac{\partial}{\partial
y'}-\beta x'\right]\varphi_{A}=\epsilon \varphi_{B}, \\
&&-i\left[\frac{\partial}{\partial x'}+i\frac{\partial}{\partial
y'}-\beta x'\right]\varphi_{B'}=\epsilon \varphi_{A'}, \\
&&-i\left[\frac{\partial}{\partial x'}-i\frac{\partial}{\partial
y'}+\beta x'\right]\varphi_{A'}=\epsilon \varphi_{B'}.
\end{eqnarray}
\end{subequations}
In the above equations, we used the following dimensionless units:
$x'=x/\sqrt{S}$, $y'=y/\sqrt{S}$, $\beta=2\pi
BS/\Phi_{0}=2\pi\Phi/\Phi_{0}$, $\epsilon=E/E_{0}$, with
$E_{0}=\hbar v_{F}/\sqrt{S}$, where $S\propto L^{2}$ is the area of
the dot with $L$ being the length of the side of the dot.
In this paper, we solve Eqs. (\ref{eqS}) numerically, using the
finite elements method, for the triangular and hexagonal graphene
flakes shown in Fig. 1, considering zigzag, armchair and infinite
mass boundary conditions. The numerical calculations are performed
by using the standard finite element package COMSOL Multiphysics
\cite{Comsol}, which discretizes the two-dimensional flake in a
finite-sized mesh and allows the implementation of the appropriate
boundary conditions. The way the boundary conditions are implemented
in the continuum model is the subject of the following three
subsections.
\subsection{Zigzag boundary conditions}
The geometry of the hexagonal and triangular graphene quantum dots
with zigzag edges are illustrated in Figs. 1(b,d). The length of
one side of the hexagonal and triangular dots, respectively, are given by
$L=\sqrt{3}(N_{s}-1/3)a$ and $L=\sqrt{3}(N_{s}+1)a$, with $N_{s}$
being the number of atoms in each side of the dot and $a=0.142$ nm
is the C-C distance. The total number of C-atoms in the triangular dot
is $N=[(N_{s}+2)^2-3]$ and $N=6N_{s}^{2}$ for the hexagonal dot. The zigzag-type
boundary condition was previously studied by Akhmerov \emph{et al.} \cite{Akhmerov}, who
presented a model which is generically applicable to any honeycomb
lattice. For a graphene dot with zigzag edges and if the last atoms
at the boundary are from sublattice A (blue circles in Fig. 1), the
boundary conditions are given by $\varphi_{A}=\varphi_{A'}=0$, whereas
$\varphi_{B}$ and $\varphi_{B'}$ are not determined, and similarly, when the
zigzag edges are terminated by the B atoms (red dots in Fig. 1),
$\varphi_{B}=\varphi_{B'}=0$, while $\varphi_{A}$ and $\varphi_{A'}$ are
not determined.

\begin{figure}
\centering
\includegraphics[width=8.5cm]{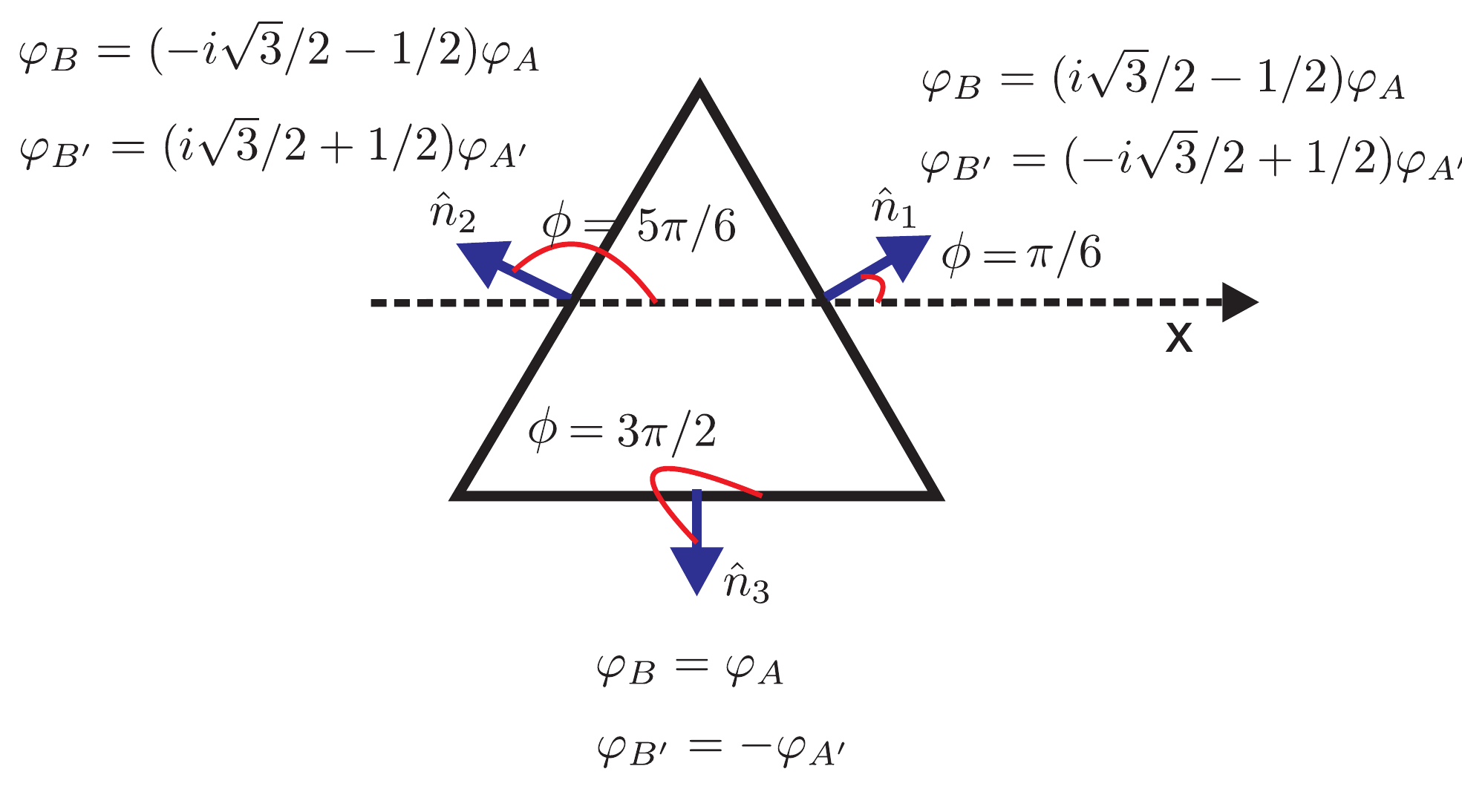}
\caption{The infinite-mass boundary conditions implemented on the
edges of a triangular dot. $\hat{n}_{1},\hat{n}_{2},\hat{n}_{3}$ are
the outward unit vectors at each edge of the dot.} \label{fignew}
\end{figure}
\begin{figure}
\centering
\includegraphics[width=8.5cm]{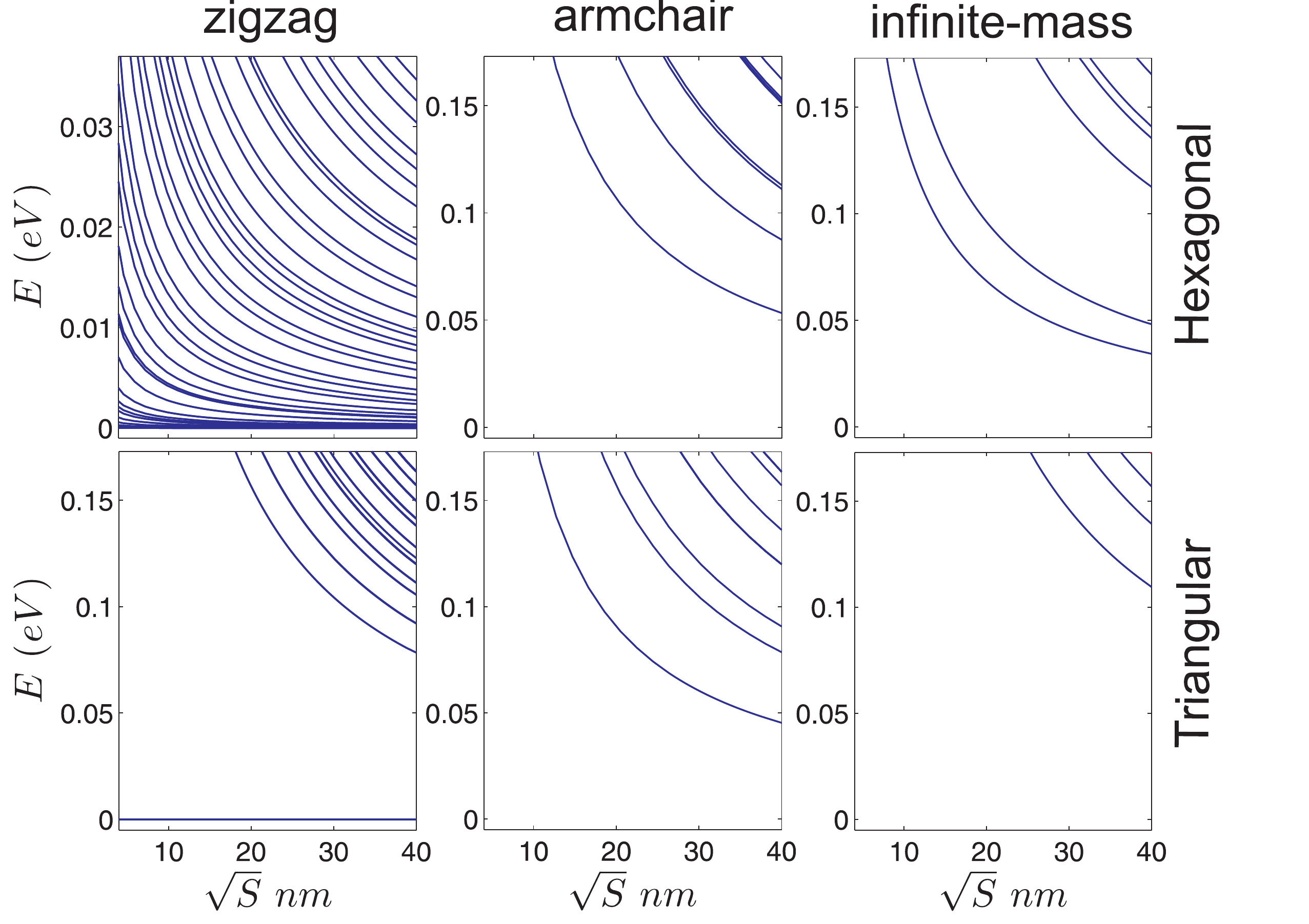}
\caption{Energy levels of hexagonal (a, c, e) and triangular (b,
d, f) graphene quantum dots with zigzag (a, b), armchair (c, d)
edges and infinite mass boundary condition (e, f) as function of
the square root of the dot area $S$ in the absence of a magnetic
field.} \label{fig1}
\end{figure}

\subsection{Armchair boundary conditions}
The geometry of a hexagonal and triangular graphene quantum dot with
armchair edges is illustrated in Figs. 1(a,c). Here, the length of
one of the edges of the hexagon dot is $L=(3N_{s}-4)a/2$ and for the
triangular dot is $L=3N_{s}a/2$. For an armchair hexagonal graphene
dot the total number of C-atoms is $N=[9N_{s}(N_{s}/2-1)+6]$ and for
the triangular dot is given by $N=(N_{s}+2)3N_{s}/4$. Note that in
the case of armchair boundaries the number of C-atoms in each side
is an even number (see Figs. \ref{fig0}(a,c)).

\begin{figure}
\centering
\includegraphics[width=8cm]{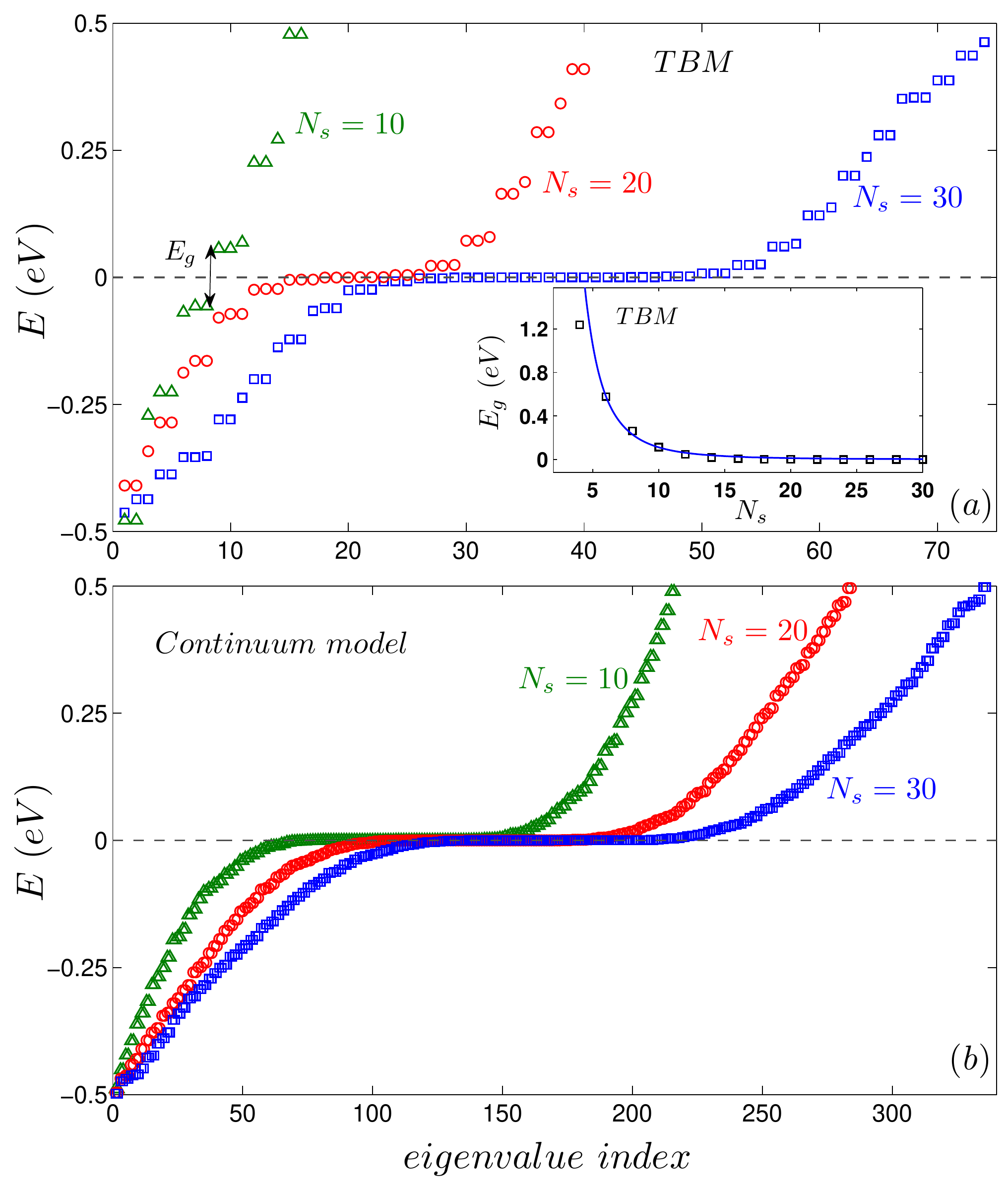}
\caption{Energy levels of a zigzag hexagonal graphene dot as
function of the eigenvalue index obtained by (a) the TBM and (b) the
continuum model, for three different sizes of the dot with $N_{s}$ =
10, 20, 30, having respectively surface area $S$ = 14.68, 60.78,
138.32 $nm^{2}$. The inset in panel (a) shows the energy gap $E_{g}$
as function of $N_{s}$ obtained, by the TBM.} \label{fig2}
\end{figure}
\begin{figure}
\centering
\includegraphics[width=8cm]{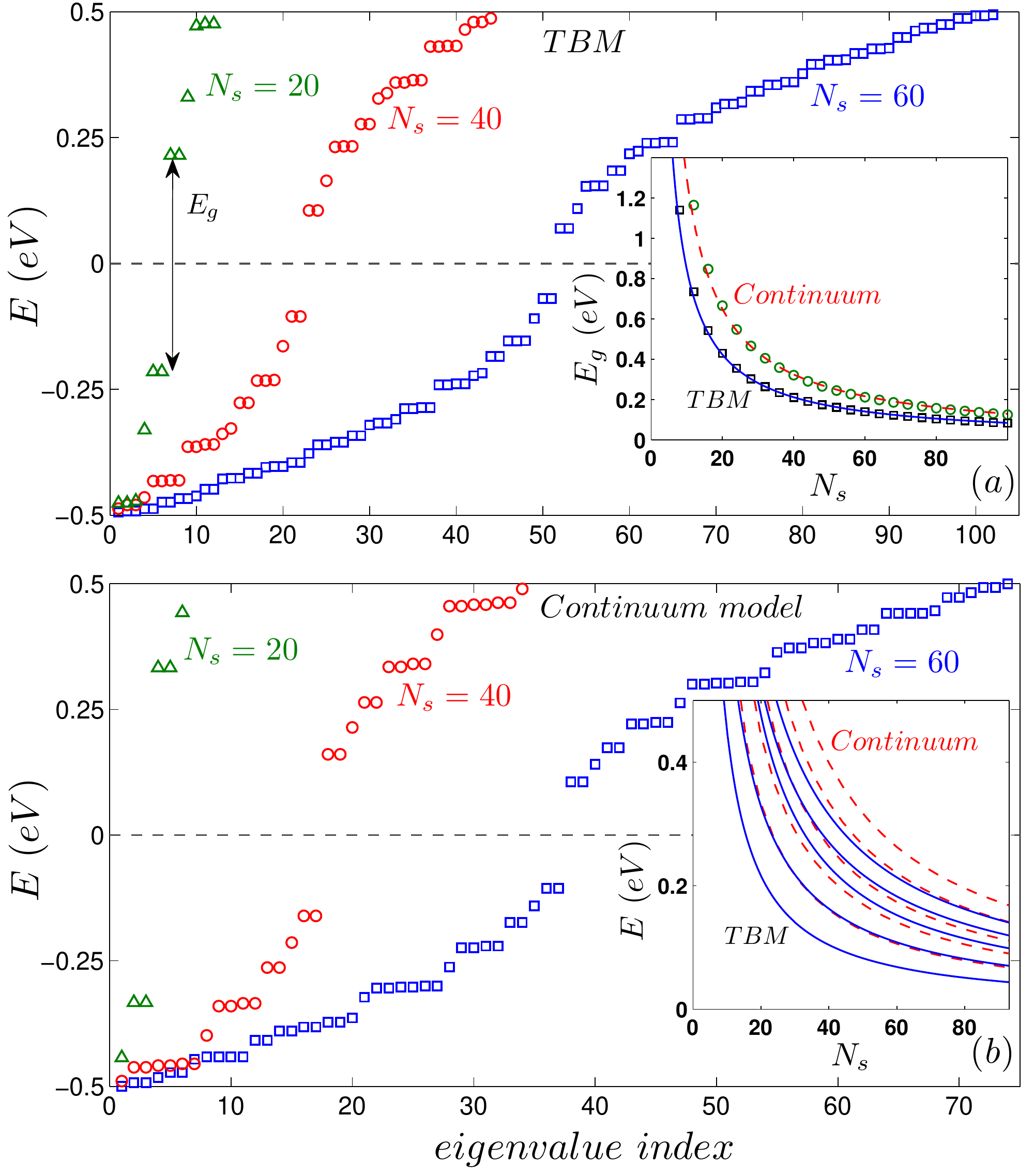}
\caption{Energy levels of an armchair hexagonal graphene dot as
function of the eigenvalue index obtained by (a) the TBM and (b) the
continuum model for three different sizes of the dot with $N_{s}$ =
20, 40, 60 having respectively surface area $S$=41.07, 176.23,
405.68 $nm^{2}$. The inset in panel (a) shows the energy gap
obtained from both TBM (black squares) and continuum model (green
circles). The inset in panel (b) shows the lowest electron energy
levels as function of $N_{s}$ for both TBM (blue solid curves) and
continuum model (red dashed curves).} \label{fig3}
\end{figure}
\begin{figure}
\centering%%%%
\includegraphics[width=8.5cm]{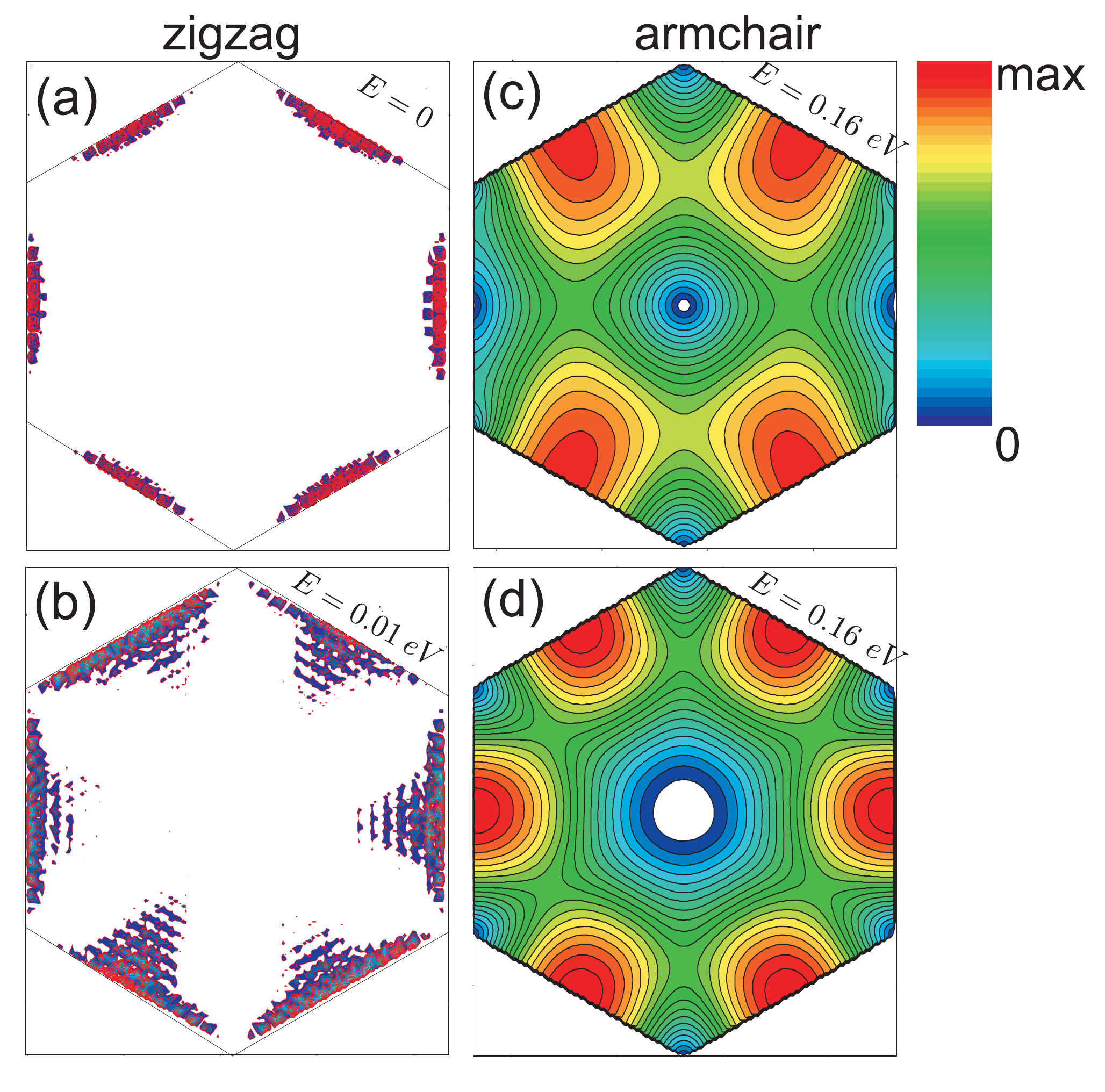}
\caption{Electron probability densities corresponding to the two
lowest energy levels of hexagonal graphene flakes, obtained by the
continuum model, for (a,b) zigzag ($N_{s}=20$) and (c,d) armchair
($N_{s}=40$) edges.} \label{WaveHexa}
\end{figure}

\begin{figure}
\centering
\includegraphics[width=8cm]{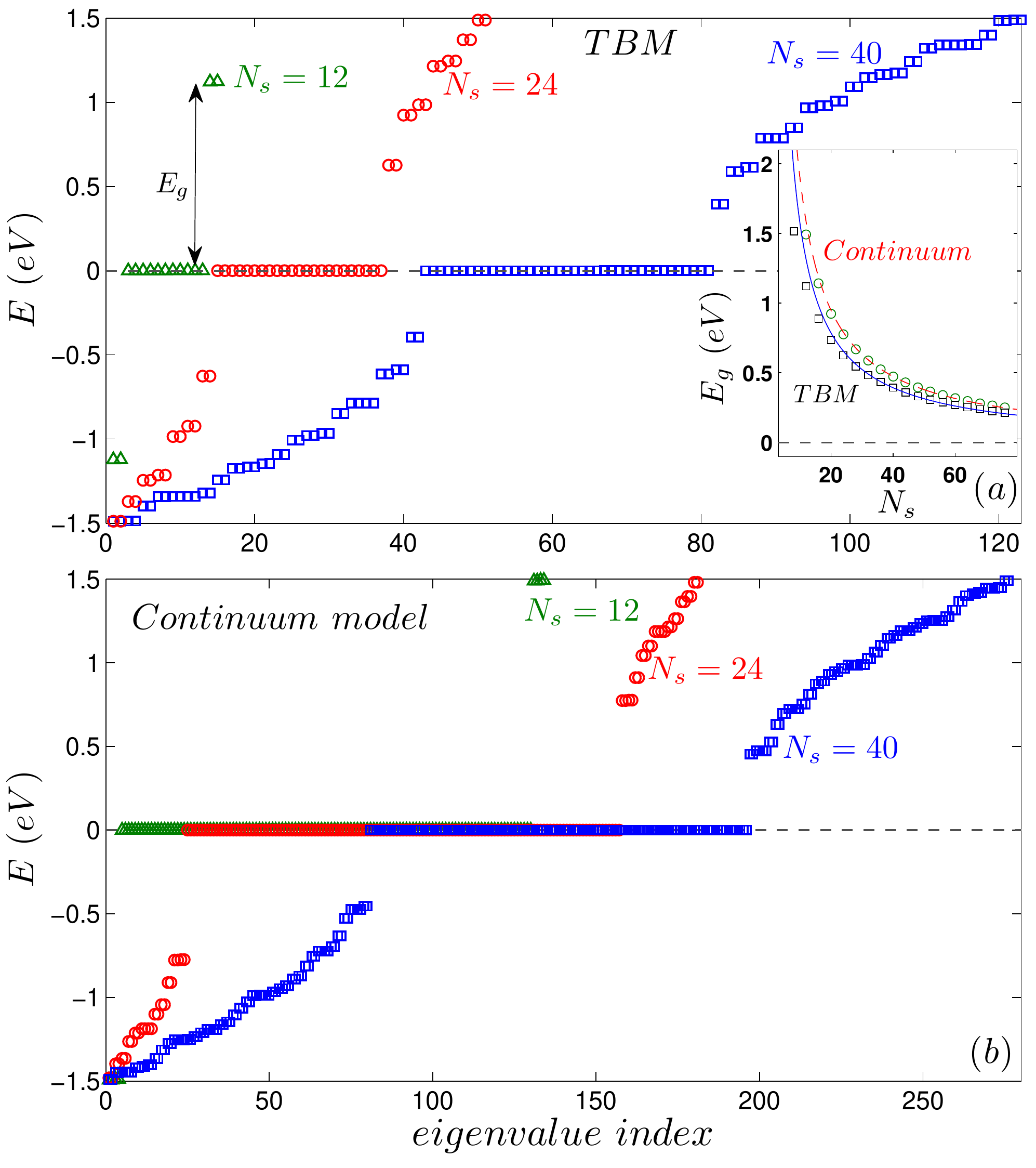}
\caption{Energy levels of a zigzag triangular graphene dot as
function of the eigenvalue index obtained by (a) the TBM and (b) the
continuum model for three different sizes of the dot with $N_{s}$ =
12, 24, 40 having respectively surface area $S$=4.42, 16.37, 44.03
$nm^{2}$. The inset in panel (a) shows the energy gap obtained from
both TBM (black squares) and continuum model (green circles).}
\label{fig4}
\end{figure}
\begin{figure}
\centering
\includegraphics[width=8cm]{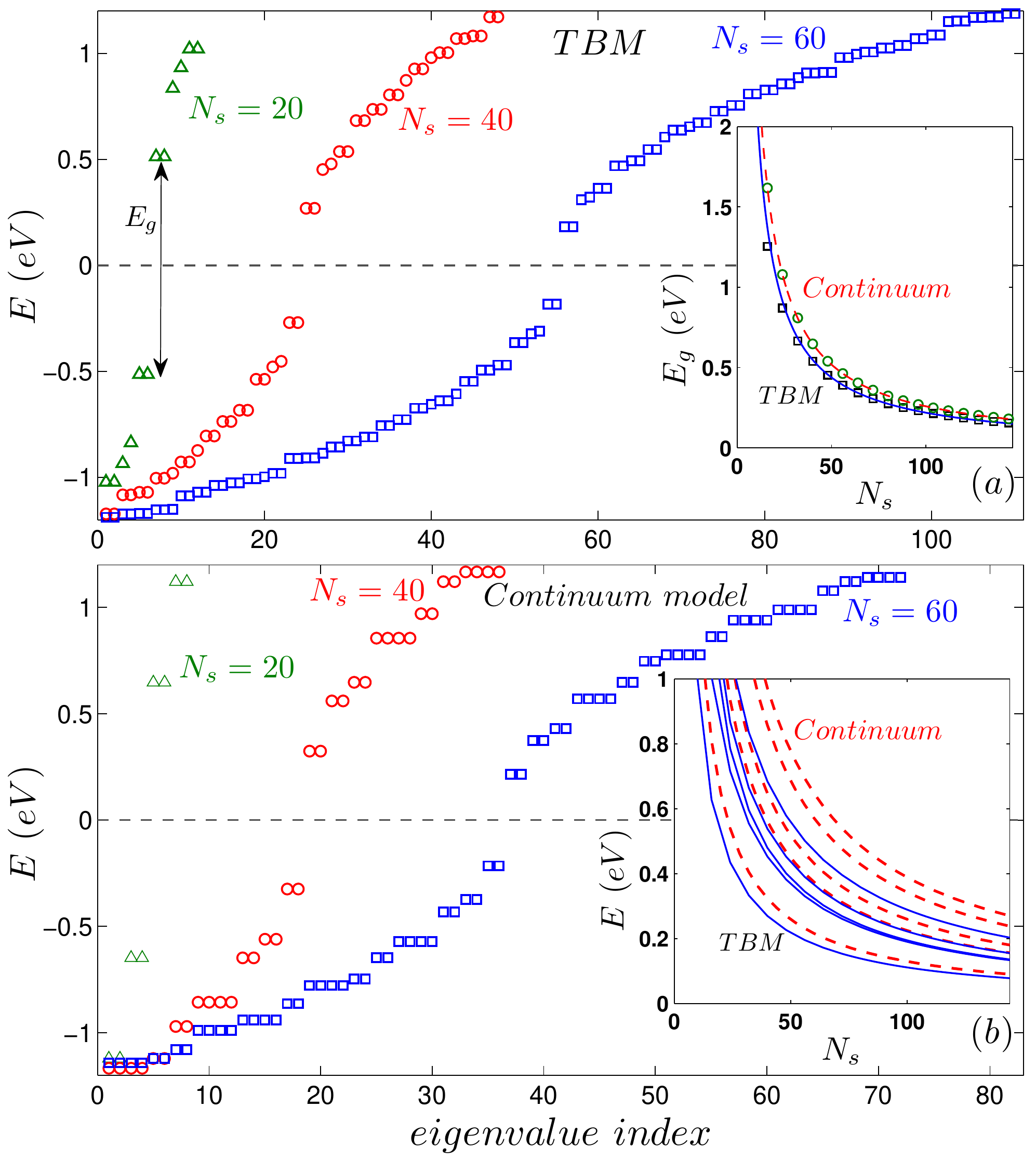}
\caption{Energy levels of an armchair triangular graphene dot as
function of the eigenvalue index obtained by (a) the TBM and (b) the
continuum model for three different sizes of the dot with $N_{s}$ =
20, 40, 60 having respectively surface area $S$=7.85, 31.43, 70.72
$nm^{2}$. The inset in panel (a) shows the energy gap obtained from
both TBM (black squares) and continuum model (green circles). The
inset in panel (b) shows the lowest electron energy levels as
function of $N_{s}$ for both TBM (blue solid curves) and continuum
model (red dashed curves).} \label{fig5}
\end{figure}
From Figs. 1(a,c), we notice that the edge atoms consist of a line
of A-B dimers, where the wave function should be zero. From Eqs.
(\ref{eqA}) and (\ref{eqB}), these boundary conditions
become\cite{Brey}
\begin{subequations}
\begin{eqnarray}\label{eq}
&&\varphi_{A}(\boldsymbol{r})=-
e^{i(\boldsymbol{K'}-\boldsymbol{K})\cdot\boldsymbol{r}}\varphi_{A'}(\boldsymbol{r}),\\
&&\varphi_{B}(\boldsymbol{r})=-
e^{i(\boldsymbol{K'}-\boldsymbol{K})\cdot\boldsymbol{r}}\varphi_{B'}(\boldsymbol{r}),
\end{eqnarray}
\end{subequations}
where $\boldsymbol{r}$ is taken at the position of the edge. Notice
that these armchair boundary conditions mix the wave functions of
the $K$ and $K'$ points.
\begin{figure}
\centering %%%
\includegraphics[width=8.5cm]{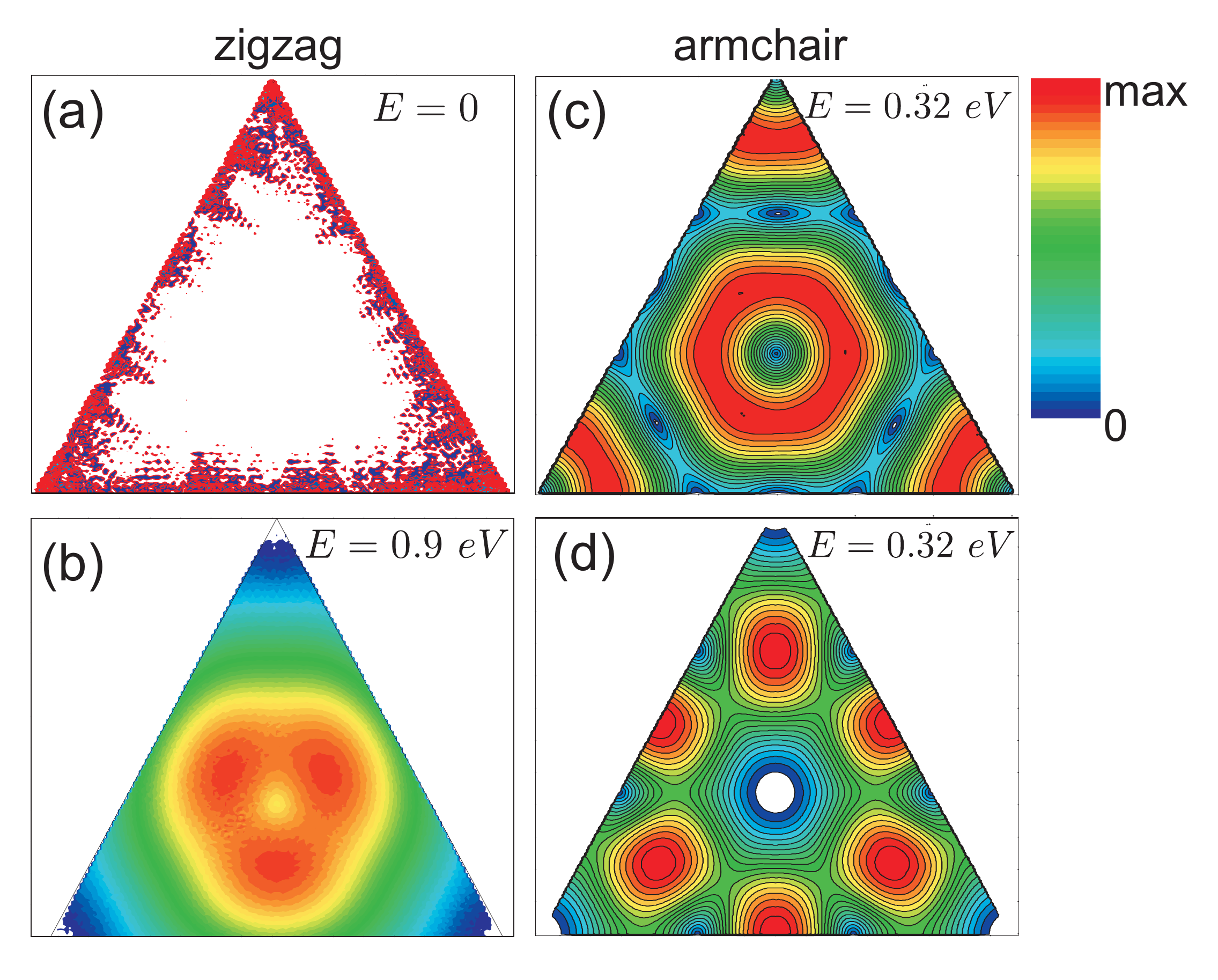}
\caption{Electron probability densities corresponding to the lowest
energy levels of the triangular graphene flakes, obtained by the
continuum model, for (a,b) zigzag ($N_{s}=20$) and (c,d) armchair
($N_{s}=40$) edges.} \label{waveTri}
\end{figure}
\begin{figure}
\centering
\includegraphics[width=9cm]{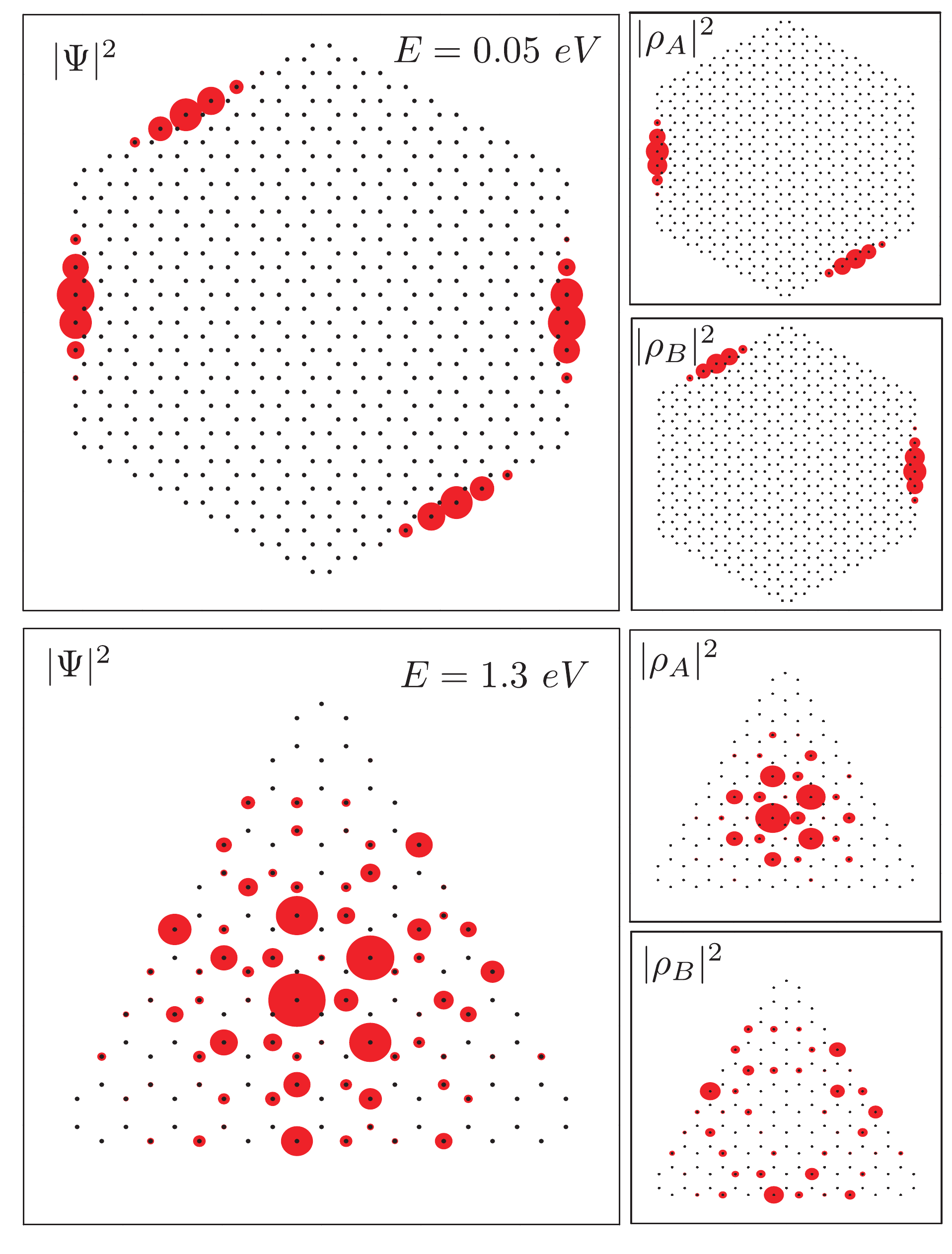}
\caption{Electron densities for the first energy level of the
triangular and hexagonal graphene flakes (using TBM) with $N_{s}=10$
and zigzag edges. Left panels show the total electron density
$|\Psi|^{2}$ and the right panels present the electron densities
associated with $A$ and $B$ sublattices. The gray dots are the
positions of C-atoms.} \label{TBMWaveZig}
\end{figure}
\begin{figure}
\centering
\includegraphics[width=9cm]{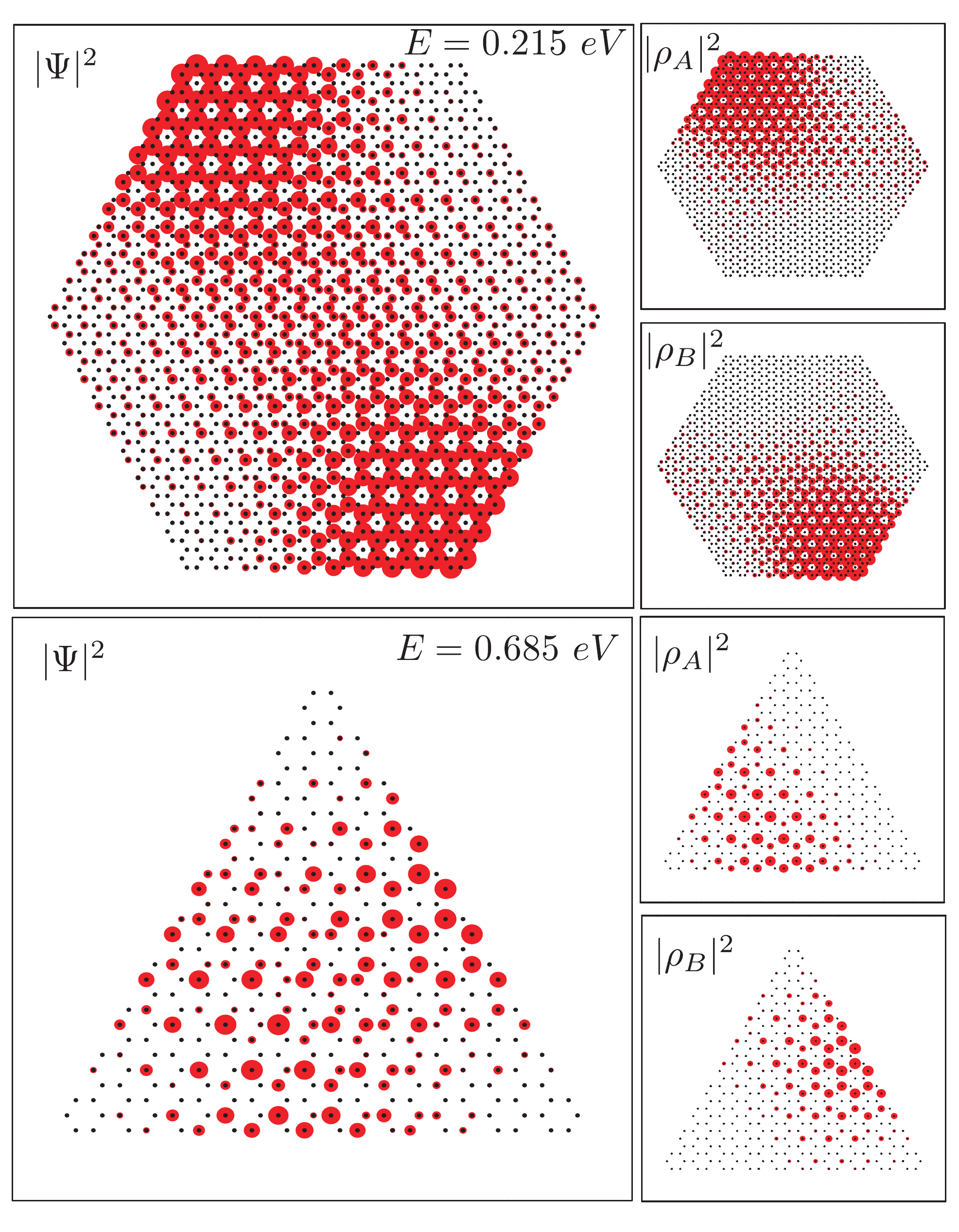}
\caption{The same as Fig. \ref{TBMWaveZig} but for the dots with armchair boundaries and $N_{s}=20$.} \label{TBMWaveArm}
\end{figure}
\subsection{Infinite-mass boundary condition}
A mass-related potential energy $V(x,y)$ can be coupled to the Hamiltonian via the
$\sigma_{z}$ Pauli matrix,
\begin{equation}\label{eq_mass}
H=v_{F}~\boldsymbol{\sigma}.~\mathbf{p}+\tau\sigma_{z}V(x,y),
\end{equation}
where the parameter $\tau=\pm1$ distinguishes the two $K$ and $K'$
valleys. It is straightforwardly verified that the presence of a
mass term in the Hamiltonian of Eq. (\ref{eq_mass}) induces a gap in
the energy spectrum of graphene. However, if the mass-related
potential $V(x,y)$ is defined as zero inside the dot and infinity at
its edge, the Klein tunneling effect at the interface between the
internal and external regions of the dot can be avoided and,
consequently, the charge carriers will be confined. This
infinite-mass boundary condition can be introduced in the Dirac
equation by defining $\varphi_{B}(x,y)/\varphi_{A}(x,y)=i e^{i\phi}$
and $\varphi_{B'}(x,y)/\varphi_{A'}(x,y)=-i e^{i\phi}$ (which,
respectively, correspond to the $K$-point and the $K'$-point
wavespinors) at the boundary, where $\phi$ is the angle between the
outward unit vector at the edges and the x-axis. \cite{Berry} Due to
its simplicity, this type of boundary condition has been used in the
study of circular graphene dots \cite{Schnez} and rings
\cite{Abergel, Recher2} in the presence of a perpendicularly
magnetic field, where analytical solutions can be found. For the
hexagonal and triangular geometries the angle $\phi$ has a fixed
value at each side of the dot that simplifies the boundary
conditions to $\varphi_{B}=\alpha\varphi_{A}$ (for the $K$ valley)
and $\varphi_{B'}=-\alpha\varphi_{A'}$ (for the $K'$ valley) where
$\alpha=ie^{i\phi}$ is a complex number. The infinite-mass boundary
conditions are shown explicitly in Fig. \ref{fignew} for a
triangular dot.
\begin{figure}
\centering
\includegraphics[width=8.5cm]{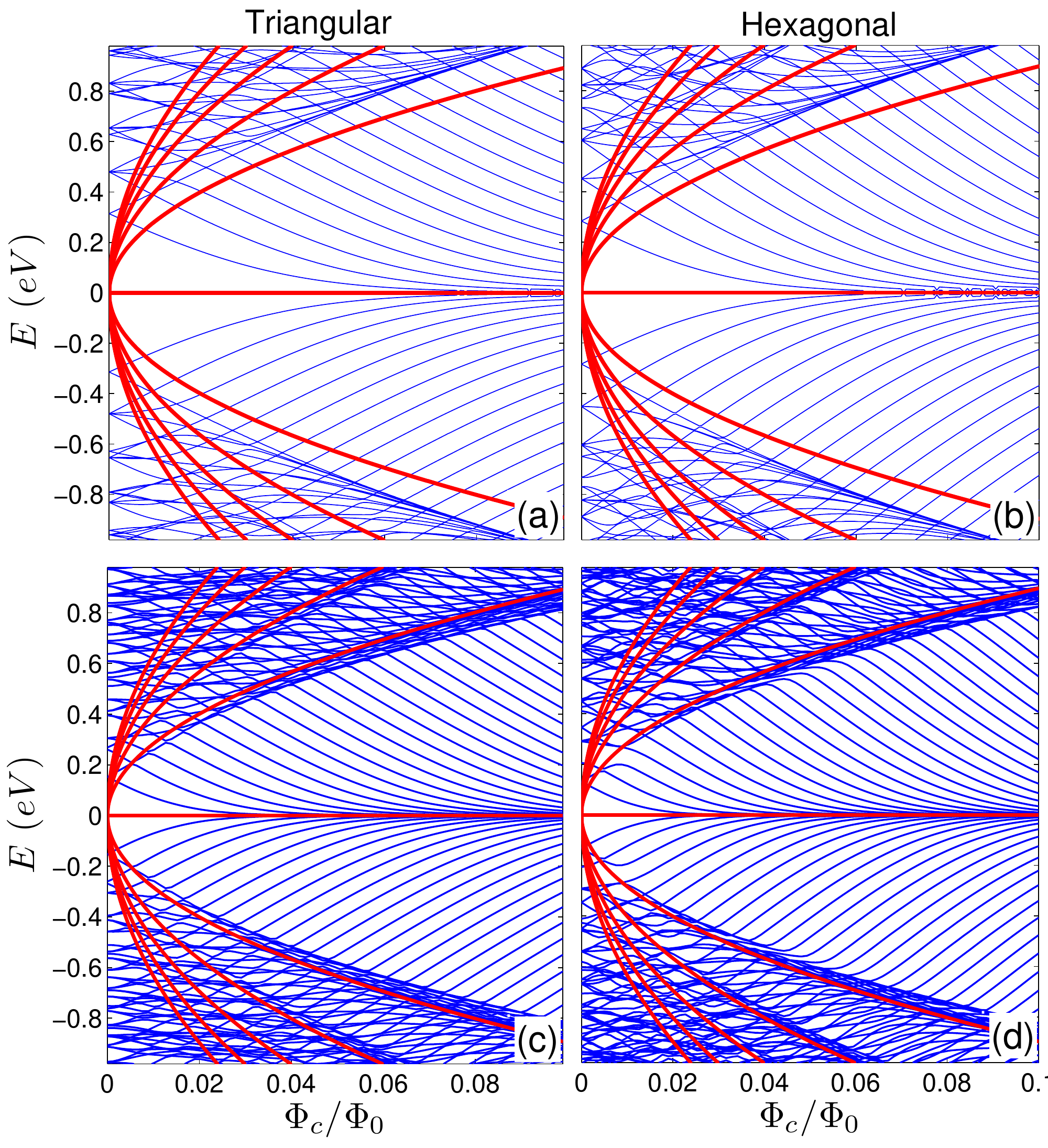}
\caption{Energy levels of (a,c) triangular and (b,d) hexagonal
graphene dots with armchair boundary as function of the magnetic
flux threading one carbon hexagon $\Phi_{c}$. The results in panels
(a,b) are obtained using the continuum model while panels (c,d)
display the TBM results. The quantum dots have an area $S$ such that
$\sqrt{S}=10$ nm.} \label{fig12}
\end{figure}
\section{Numerical results}
\subsection{Zero magnetic field}
The energy levels of hexagonal (upper panels) and triangular (lower
panels) graphene flakes, as calculated within the continuum model,
are shown in Fig. \ref{fig1} as function of the square root of the
dot area. The results are shown for zigzag (a,b), armchair (c,d) and
infinite-mass (e,f) boundary conditions and are qualitatively and
quantitatively very different. As the dot area increases, the energy
levels tend to a gapless spectrum, which is expected, since the
energy spectrum of an infinite graphene sheet does not exhibit a
gap. A peculiar spectrum is observed for zigzag triangular dots
(Fig. \ref{fig1}(b)): zero energy states are found for all sizes of
such a dot. These zero energy states are separated from the
remaining positive and negative energy states by an energy gap which
decreases as the dot becomes larger. The presence of such zero
energy states in triangular and trapezoidal graphene flakes have
been previously reported in the literature \cite{Guclu, Heiskanen,
Akola}, where the TBM was applied. In the case of zigzag triangular
dots, it has been shown analytically\cite{Guclu} that the equation
$H\Psi = 0$ for the TBM Hamiltonian in Eq. (\ref{eq_ham_TB2}) leads
to $N_s - 1$ linearly independent states, namely, $N_s - 1$
degenerate states with $E = 0$, for any number $N_s$ of C-atoms in
one of the edges of the flake. Thus, Fig. \ref{fig1}(b) demonstrates
that the existence of zero energy states, which is observed in the
TBM, is qualitatively captured by the approximations of the
continuum model as well. The results in Fig. \ref{fig1} also show
that the energy levels for a dot with armchair and infinite-mass
boundary conditions are qualitative more similar to each other than
the spectra for zigzag edges, where carriers are predominantly
confined at the edge of the dot. In fact, for the triangular
geometry, the infinite mass boundary condition describes very well
the armchair states, specially for lower energy states. However, for
the hexagonal geometry, the results for armchair and infinite mass
boundary conditions are only qualitatively similar where, the
hexagonal dots with infinite mass boundary condition exhibit more
energy states in comparison with the armchair case.
\begin{figure}
\centering
\includegraphics[width=8.5cm]{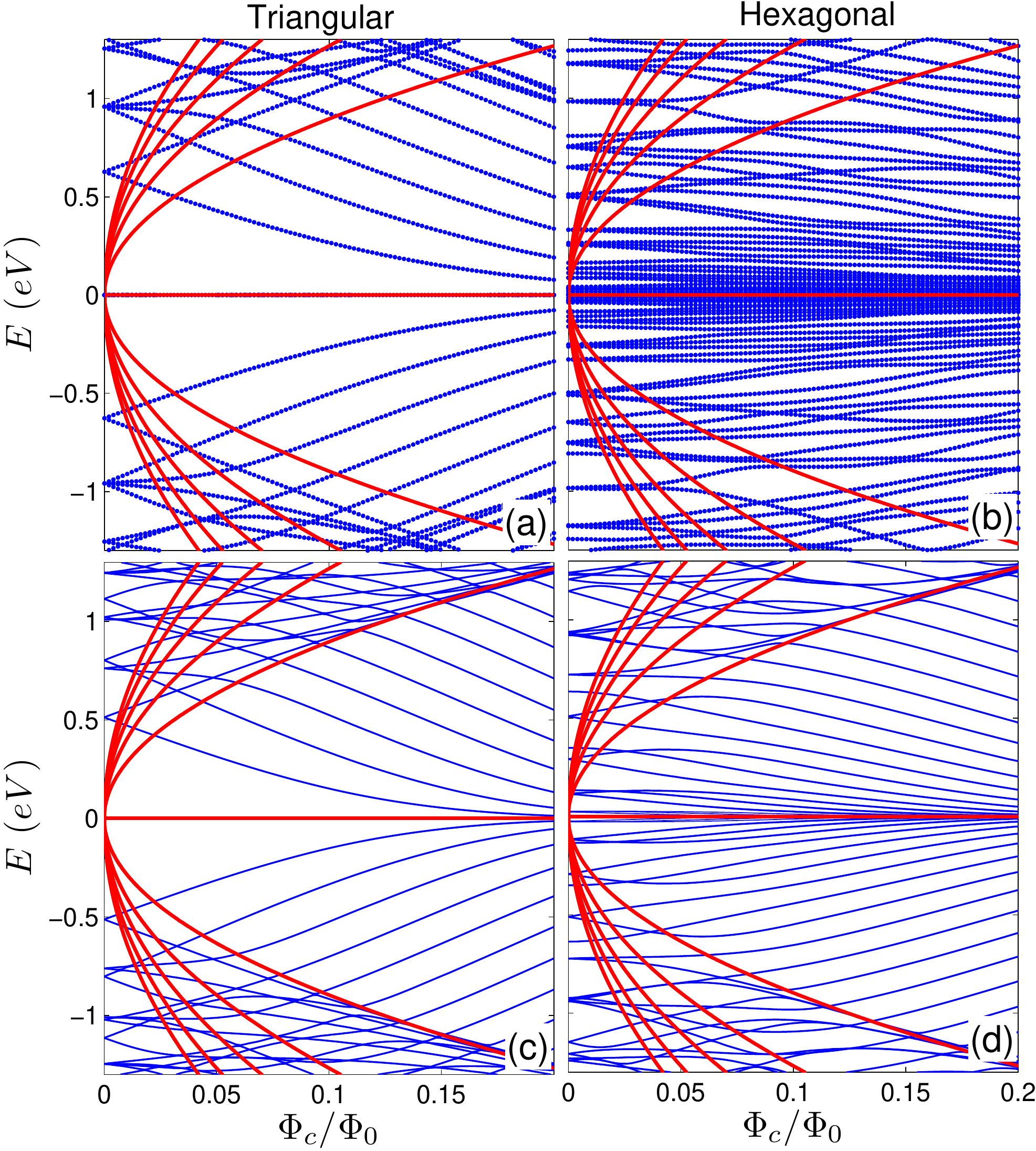} %
\caption{The same as Fig. \ref{fig12} but for dots with zigzag
boundaries and the dots have an area $S$ such that $\sqrt{S}=5$ nm.}
\label{fig13}
\end{figure}

Notice that the energy spectra shown in Fig. \ref{fig1} exhibits
degenerate states. These degeneracies, which will be evidenced in
the following figures, where we plot the energy spectra as a
function of the eigenvalue index, are related to the symmetries of
the triangular and hexagonal dots, as we will explain in further
detail later on, when we discuss about the electron probability
densities.

A comparison between the energy spectra obtained by means of the TBM
(a) and the Dirac equation (b) for zigzag hexagonal dots is shown in
Fig. \ref{fig2}, for three sizes of the dot, defined by the number
of C-atoms in each side of the hexagon $N_{s}$. The energies $E_i$
are plotted as a function of the eigenvalue index $i$. Although the
results are quantitatively different, they are qualitatively
similar, e.g. as the size of the dot increases, they start to
exhibit an almost flat energy spectrum as a function of the
eigenvalue index around the Dirac point. Such a flat spectrum leads
to a peak in the DOS close to the Dirac point, which was recently
reported in the literature \cite{Zhang} for graphene dots with
zigzag edges within the TBM. The curves for $N_s = 30$ obtained by
the TBM and continuum models are very similar, except for the fact
that many more states are found in the latter, whereas the discrete
character of the spectrum in the former is much more clear. For
smaller dots, the agreement between these two models becomes clearly
worse. For instance, an energy gap $E_{g}$ is found for very small
hexagons (i.e. $N_s\leq10$) within TBM, whereas in the case of the
continuum model such a gap is extremely small. As a consequence, the
continuum model overestimates the DOS at $E = 0$ as the dot size
decreases, since it exhibits a plateau in the energy as a function
of the eigenstate index in the vicinity of $E = 0$ even for smaller
$N_s$, where TBM results show a gap in the energy spectrum. Notice
that the $E = 0$ states in zigzag dots are edge states, so that the
number of zero-energy states depends on the number of edge atoms in
the TBM and, similarly, to the number of mesh elements at the edge
in the continuum model. Therefore, in the continuum model for $E=0$,
the finite elements problem is ill-defined, where the constructed
matrix of the finite mesh elements in this case is singular (zero
inverse), leading to spurious solutions around $E = 0$. As the size
of the dot increases, the gap in the TBM results quickly reduces to
zero and a zero energy level for the hexagonal flakes with zigzag
edges appears.\cite{Peres} In the inset of Fig. \ref{fig2}(a), the
energy gap values obtained by TBM are shown as function of $N_{s}$.
These results can be fitted to $E_{g}=\alpha(1/N_{s})^{\gamma}$
(blue solid curve in the inset of Fig. \ref{fig2}(a)), where
$\alpha=94.6$ eV and $\gamma=3.23$ are fitting parameters.

\begin{figure}
\centering
\includegraphics[width=8.5cm]{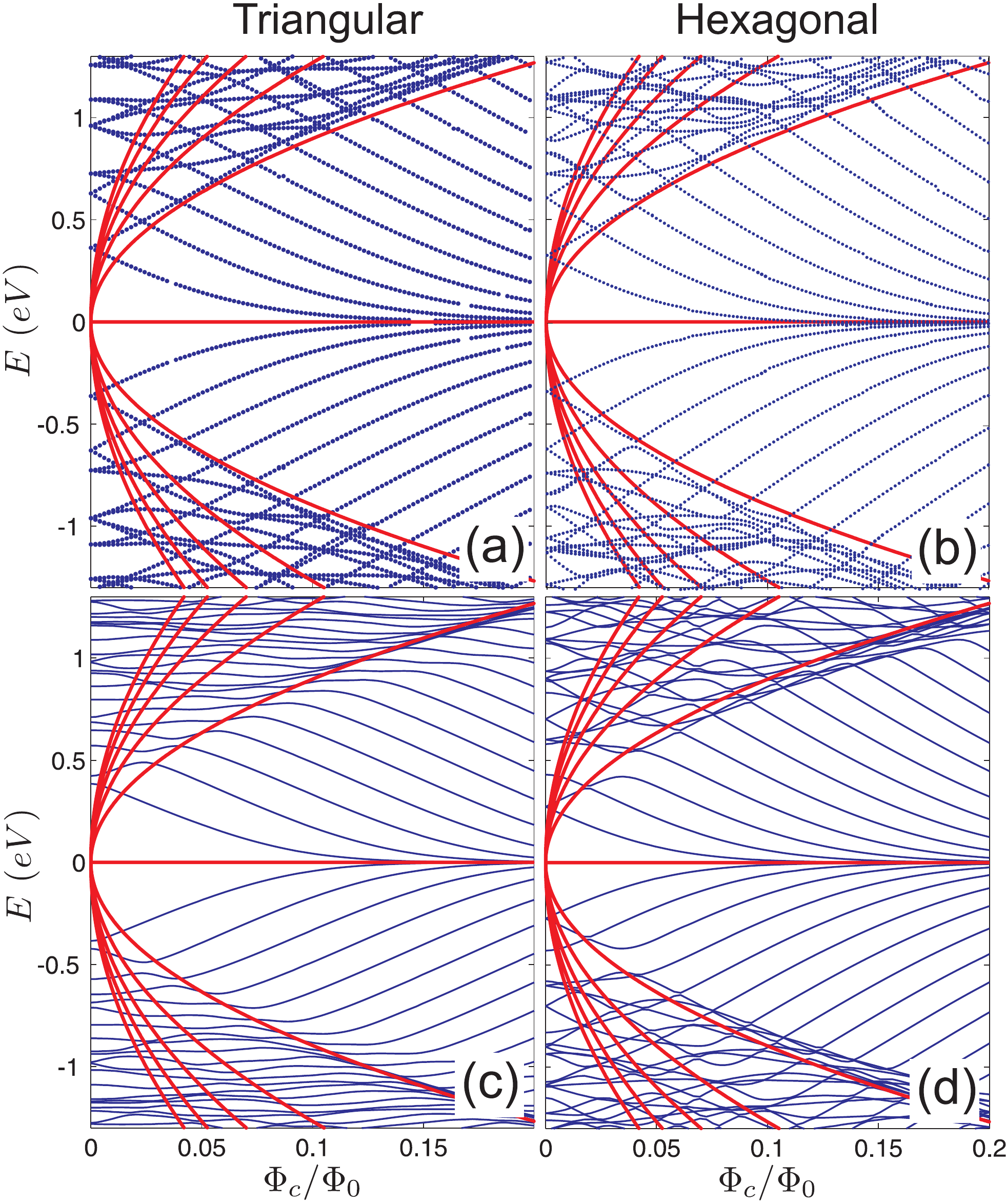}
\caption{The same as Fig. \ref{fig12}, but for dots with
infinite-mass boundary conditions. The dots have an area $S$ such
that $\sqrt{S}=5$ nm. In panels (c,d), the infinite-mass boundary
condition is applied within the TBM model, where we imposed a +10(-10)
eV on-site potential for sublattice A(B) around the dot geometry.}
\label{fig14}
\end{figure}

The energy states of armchair hexagonal dots are shown as a function
of the eigenvalue index in Fig. \ref{fig3} within the TBM approach
(Fig. \ref{fig3}(a)) and the Dirac-Weyl equations (Fig.
\ref{fig3}(b)), for three different sizes of the dot. The energy
spectrum in both cases approach the prolonged S-shape curve
predicted by Ezawa\cite{Ezawa} as the size of the dot increases and
the spectrum exhibits an energy gap $E_{g}$ at the Dirac point. The
energy gap as function of $N_{s}$ is shown in the inset of Fig.
\ref{fig3}(a) which decreases rapidly as the size of the dot
increases. Our numerical results can be fitted to
$E_{g}=\alpha/N_{s}$ with $\alpha=8.5$ eV for the TBM (blue solid
curve) and $\alpha=13$ eV for the continuum model (red dashed curve)
results. Notice that $E_{g}$ obtained from the continuum model is
larger than the one from the TBM results in particular for small
$N_{s}$ and both curves can not be made to coincide by a simple
shift in $N_{s}$. This is clearly a consequence of the increased
importance of corrections to the linear spectrum used in the
continuum model for small sizes of the system. The inset of Fig.
\ref{fig3}(b) shows the five lowest electron states for both TBM
(blue solid curves) and the continuum model (dashed red curves). Our
results show that the continuum model overestimates the energy
values also for the upper energy levels in comparison with the TBM
energy levels. In fact, the energy dispersion in the continuum model
is given by a linear curve, which coincides with the TBM energy
spectrum for low energies, but as the energy goes further away from
$E = 0$, this linear dispersion overestimates the energy as compared
to the real band structure of graphene, which starts to bend down
from the linear spectrum as the energy increases. This emphasizes
once again the importance of the higher order corrections to the
linear dispersion, especially for high energy states and smaller dot
sizes.

Figure \ref{WaveHexa} shows the probability density (using the
continuum model) corresponding to the first two energy levels of
hexagonal flakes. The probability density for the zigzag case with
$N_{s}=20$ is presented in panels (a) and (b), respectively, for
$E=0$ and $E=0.01$ eV. The results clearly demonstrate that the zero
energy states in the zigzag case are due to edge effects and,
accordingly, are confined at the edges while the carriers confine
towards the center of the flake with increasing energy (see Fig.
\ref{WaveHexa}(b)). The probability densities of the armchair edged
graphene flake with $N_{s}=40$ are very different as seen in Figs.
\ref{WaveHexa}(c,d) for the lowest degenerate states with
$E=0.16$ eV. The electron wavefunction is spread out over the whole
sample, but different from the usual quantum dots with parabolic
energy-momentum spectrum, it has a local minimum in the center of
the dot. Note that Fig. \ref{WaveHexa}(c) has only two-fold symmetry
while Fig. \ref{WaveHexa}(d) is six-fold symmetric. Both densities
are zero in the center, while Fig. \ref{WaveHexa}(c) has two extra
zeros at the sides along $y=0$. These results are comparable to the
TBM results obtained in Ref.\onlinecite{Zhang}.
\begin{figure}
\centering
\includegraphics[width=8cm]{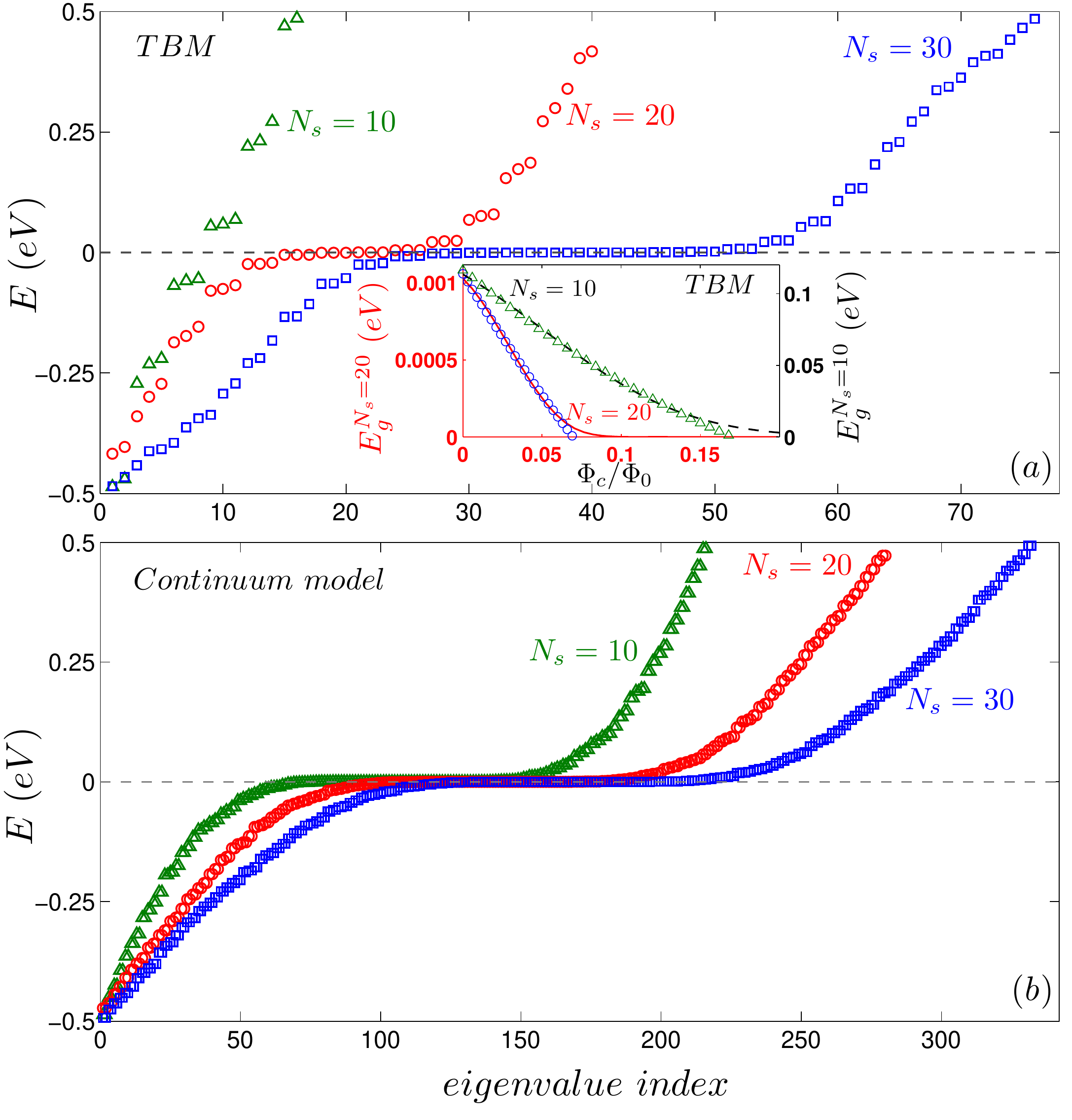}
\caption{The same as Fig. 3 but in the presence of an external
magnetic field of $B=50$ T. The inset shows the energy gap as
function of the magnetic flux obtained by the TBM for two values of
$N_{s}$. The triangle and circle symbols display Eq. (13) which is
fitted to the numerical results.} \label{fig8}
\end{figure}

The energy spectrum for triangular dots with zigzag edges, obtained
by the TBM and the Dirac-Weyl equation are shown as a function of
the eigenvalue index in Figs. \ref{fig4}(a) and \ref{fig4}(b),
respectively. Notice that both energy spectra exhibit zero energy
states. As we mentioned before, the number of degenerate states with
zero energy is a well defined quantity in the tight-binding
approach, namely, $N_{s}-1$, where $N_{s}$ is the number of C-atoms
in one side of the triangle \cite{Guclu}. On the other hand, the
result in Fig. \ref{fig4}(b) for the continuum model exhibits many
more zero energy states. Therefore, while the continuum model
captures qualitatively the existence of zero energy states, it does
not provide the appropriate number of degenerate states as
calculated by the TBM. These zero energy levels are related to the
edge states of zigzag graphene flakes \cite{Zhang, Guclu}. The
energy gap (between the zero energy level and the first non-zero
eigenvalue) is shown in the inset of Fig. \ref{fig4}(a) as function
of the size of the dot, where $E_{g}$ obtained by both models are
comparable and the difference between the TBM (red dashed curve) and
continuum (blue solid curve) results tends to zero for large
graphene flakes. These results can be fitted to $E_{g}=\alpha/N_{s}$
with $\alpha=15.75$ eV for the TBM gap and $\alpha=18.9$ eV for the
continuum model.

\begin{figure}
\centering
\includegraphics[width=8cm]{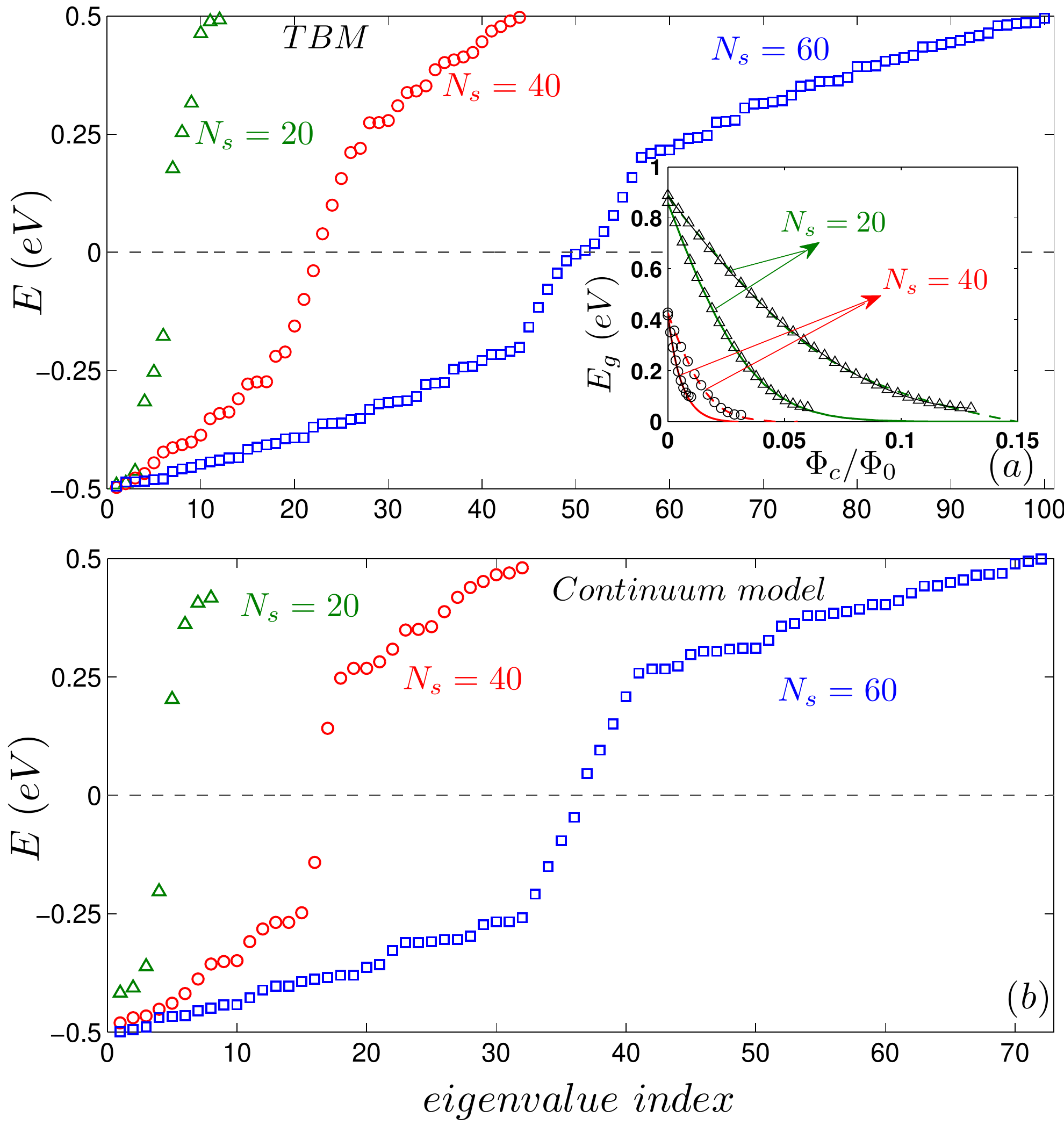}
\caption{The same as Fig. 4 but in the presence of an external
magnetic field of $B=50$ T. The inset shows the energy gap as
function of the magnetic flux obtained by the TBM (solid curves) and
continuum model (dashed curves) for two values of $N_{s}$. The
triangle and circle symbols display Eq. (13) which is fitted to the
numerical results.} \label{fig9}
\end{figure}

The energy spectra of triangular dots with armchair edges obtained
by the TBM and the continuum model are shown in Fig. \ref{fig5}. No
zero energy states are found and the energy gap at the Dirac point
for both models is comparable. The gap can be fitted to
$E_{g}=\alpha/N_{s}$ ($\alpha=21.9$ eV for TBM and $\alpha=25.9$ eV
for the continuum model) as shown respectively by the blue solid and
dashed red curves in the inset of Fig. \ref{fig5}(a). The lowest
electron energy levels, obtained by the TBM (blue solid curves) and
the continuum model (red dashed curves), are shown in the inset of
Fig. \ref{fig5}(b) as function of $N_{s}$. The results show a larger
difference between the TBM and continuum energy values for the upper
energy levels (e.g. $|E^{T}_{1}-E^{C}_{1}|<|E^{T}_{2}-E^{C}_{2}|$).

Notice that the energy gaps found for all the systems that we
investigated were fitted to $E_g = \alpha/N_s$ for different values
of $\alpha$, except for the case of zigzag hexagonal dots, where the
gap is fitted to $E_g = \alpha/N_s^{\gamma}$, with $\gamma = 3.23$.
This is a consequence of the fact that the corners of the zigzag
hexagonal dot structure are not terminated by a single atom, as in
the case of zigzag triangular dots, but by a pair of C-atoms
corresponding to two different sublattices, forming a A-B dimer (see
Fig. 1). These A-B dimers are responsible for a vanishing wave
function in the corners of the zigzag hexagonal dots, as observed in
Fig. \ref{WaveHexa}. As explained in Sec. III A, the zigzag boundary
condition for each side of the dot is implemented in the Dirac-Weyl
equations by setting to zero the component of the pseudo-spinor
corresponding to the sublattice that forms that side. As the
sublattice types of adjacent sides of a zigzag hexagonal dot are
different, connected by the A-B dimers in the corners, the whole
wave function must vanish at these corners, since these points are
composed of both A and B sublattices. The vanishing wave function at
the corners reduce the effective confinement area and, consequently,
increases the energy gap, especially for smaller dots, where the
influence of the corners is more significant. As the size of the dot
increases, the role of the corners in the energy gap becomes less
important and is eventually suppressed by the influence of the
zigzag edges, leading to the zero energy states that form the
plateau in Fig. \ref{fig2}, explaining the faster decay of the
energy gap ($\gamma = 3.23$) in zigzag hexagonal dots, as compared
to the other cases ($\gamma = 1$).

The probability density corresponding to the first two energy levels
of triangular graphene flakes, obtained by the continuum model, is
shown in Fig. \ref{waveTri}. The probability density for the zigzag
edged dot with $N_{s}=20$ is presented in panels (a) and (b),
respectively, for $E=0$ and for the first non-zero eigenvalue (i.e.
$E=0.92$ eV). For the degenerate zero energy states the carriers are
confined at the edges of the triangular flake which is typical for
zigzag boundaries. States corresponding to large energy values are
confined in the center of the triangle (Fig. \ref{waveTri}(b)). For
armchair triangular flakes, as in the hexagonal case, the electron
state is spread out over the whole flake (Figs. \ref{waveTri}(c,d)
display the different probability densities for $N_{s}=40$
corresponding to the first degenerate eigenvalues with $E=0.32$ eV).
Both wavefunctions have three-fold symmetry and the inner part is
even six-fold symmetric. Note that the electron density in Fig.
\ref{waveTri}(d) is zero at the three corners and in the center of
the triangle which is different from Fig. \ref{waveTri}(c) where
zero's are found at the corners of the inner hexagon and at the
center of the sides of this hexagon.

The TBM electron densities of the zigzag graphene dots with
$N_{s}=10$ is shown in Fig. \ref{TBMWaveZig} for the first energy
level of the triangular and hexagonal graphene flake. Left panels
present the total electron density $|\Psi^{2}|$ and the electron
densities associated with $A$ and $B$ sublattices
($|\rho_{A,B}|^{2}$) are shown in the right panels. We found that
the wavefunctions of the two-fold degenerate states are related to
each other by a $60^{\circ}$ rotation. The sum of the densities of
the degenerate states results in a six-fold (three-fold) symmetric
wavefunction for the hexagonal (triangular) flakes. As seen in Fig.
\ref{TBMWaveZig} the total electron density is related to the
densities of $A$ and $B$ sublattices by
$|\Psi|^{2}=|\rho_{A}|^{2}+|\rho_{B}|^{2}$. Figure \ref{TBMWaveArm}
describes the density distributions of the lowest energy levels for
armchair graphene flakes. For the armchair hexagonal dots the
electron densities corresponding to the $A$ and $B$ sublattices
(right panels) can be transformed to each other by a $180^{\circ}$
rotation whereas the density of the triangular wavespinors can not
be linked to each other by a rotational transformation.
\begin{figure}
\centering
\includegraphics[width=8cm]{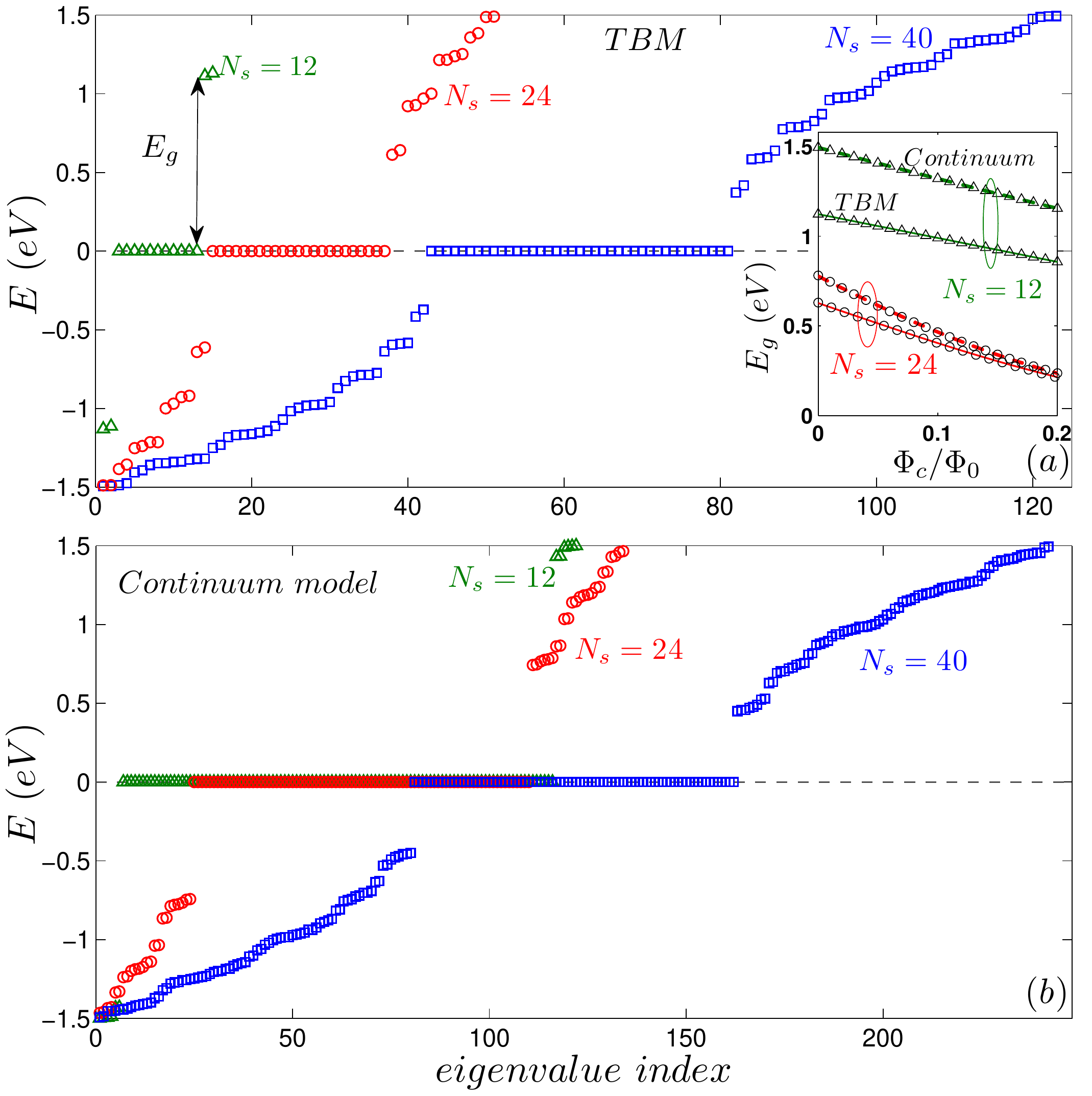}
\caption{The same as Fig. 5 but in the presence of an external
magnetic field of $B=50$ T. The inset shows the energy gap as
function of the magnetic flux obtained by the TBM (solid curves) and
continuum model (dashed curves) for two values of $N_{s}$. The
triangle and circle symbols display Eq. (13) which is fitted to the
numerical results.} \label{fig10}
\end{figure}
\subsection{Magnetic field dependence}
The dependence of the energy levels of triangular (a) and hexagonal
(b) graphene flakes on the magnetic flux through one carbon hexagon
$\Phi_{c} = BS_{c}$ is shown in Figs. \ref{fig12} and \ref{fig13},
respectively for flakes with armchair and zigzag edges. The results
in panels (a),(b) are obtained using the continuum model and the
results in panels (c),(d) show the TBM energy spectrum.
$S_{c}=(3\sqrt{3}/2)a^{2}$ is the area of a carbon hexagon which is
indicated by the yellow region in Fig. \ref{fig0}(a). The results
are obtained for dots with an area of $S = 100$ nm$^2$ and $S=25$
nm$^2$ respectively for armchair and zigzag edges. The continuum and
TBM results are qualitatively similar to each other in the sense
that as the magnetic flux increases, the energy levels converge to
the Landau levels of a graphene sheet $E_{n}$ (see red solid
curves), which are given by
\begin{equation}
E_{n}=sgn(n)\frac{3at}{2l_{B}}\sqrt{2|n|},
\end{equation}
where, $l_{B}=\sqrt{\hbar/eB}$ is the magnetic length and $n$ is an
integer. The interplay between the quantum dot confinement and the
magnetic field confinement is responsible for the appearance of a
series of (anti)-crossings in the energy spectrum. As explained
earlier, armchair graphene dots do not exhibit zero energy states
for $B=0$. However, as the magnetic field increases, some of the
excited energy levels approach the zero energy Landau level $n = 0$
in both armchair and zigzag graphene flakes, which naturally
produces (anti)-crossings between the excited states. Lifting the
degeneracy of the energy levels by the magnetic field results in a
closing of the energy gap with increasing magnetic field. Notice
that the zero energy states of zigzag triangular dots (Fig.
\ref{fig13}(a)) are not affected by the magnetic field because they
are strongly confined at the edges of the dot. All these features
are qualitatively similar to those obtained by the TBM (see the
lower panels in Figs. \ref{fig12},\ref{fig13}). In the case of
hexagonal zigzag graphene dots (see Fig. \ref{fig13}(b)), the
continuum model exhibits a plethora of additional lines as compared
to the well known energy spectrum obtained by the TBM (compare Figs.
\ref{fig13}(b) and \ref{fig13}(c)).

For the infinite-mass boundary condition, the energy spectrum of
triangular (a) and hexagonal (b) dots as a function of the magnetic
field is shown in Fig. \ref{fig14} for the dot with area $S=25$
nm$^{2}$. The energy spectrum in this case differs from both
obtained for zigzag and armchair boundary conditions. The spectra
exhibit no zero energy state at $B = 0$ and show crossings and
anti-crossings between the higher energy levels which resemble the
TBM results (see Figs. \ref{fig14}(c),(d) respectively for
triangular and hexagonal dots). In the TBM model, the infinite-mass
boundary conditions can be realized as a graphene dot structure
surrounded by an infinite mass media, where we applied a staggered
potential (i.e. +10(-10) eV on-site potential for sublattice A(B))
around the dot geometry.

The energy levels obtained by the TBM (a) and the continuum model
(b) for hexagonal graphene flakes under a $B = 50$ T external
magnetic field are shown in Figs. \ref{fig8} and \ref{fig9} for
zigzag and armchair edges, respectively, as function of the
eigenvalue index. The energy spectra of such systems in the absence
of magnetic field, which are shown in Figs. \ref{fig2} and
\ref{fig3}, are composed of a series of degenerate states for $|E| >
0$. The magnetic field lifts the degeneracy of such states and
reduces the gap between the states. The energy gap as function of
the magnetic flux through a single carbon hexagon $\Phi_{c}$ is
shown in the insets of Fig. \ref{fig8}(a) and Fig. \ref{fig9}(a)
respectively for zigzag (with $N_{s}=10,20$) and armchair (with
$N_{s}=20,40$) hexagonal dots. These results can be fitted to
\begin{equation}
E_{g}(\Phi_{c}/\Phi_{0})=E_{g}^{0}+E_{g}^{1}(\Phi_{c}/\Phi_{0})+E_{g}^{2}(\Phi_{c}/\Phi_{0})^2
\end{equation}
where $E_{g}^{0,1,2}$ (eV) are the fitting parameters. In the inset
of Figs. \ref{fig8}(a) and \ref{fig9}(a), the fitted results are
shown by symbols. The fitting parameters for the TBM results of a
zigzag hexagonal dot with $N_{s}=10$ (for the range of
$0\leq\Phi_{c}/\Phi_{0}\leq 0.17$ ) are
$E_{g}^{0,1,2}=(0.12,-0.91,1.36)$ eV (see triangles in the inset of
Fig. \ref{fig8}(a)) and $E_{g}^{0,1,2}=(0.86,-26,210)$ eV,
$E_{g}^{0,1,2}=(0.88,-12.5,46.5)$ eV are the fitting parameters of
an armchair hexagonal dot with $N_{s}=20$ respectively for TBM and
continuum results (triangles in the inset of Fig. \ref{fig9}(a)).
The fittings are done for the range of $0\leq\Phi_{c}/\Phi_{0}\leq
0.06$ and $0\leq\Phi_{c}/\Phi_{0}\leq 0.13$ respectively for TBM and
continuum results.

For the zigzag case and for $N_s = 20$, the energy gap is already
negligible, whereas for $N_s = 10$, the $E_g \approx 0.12$ eV gap at
$B = 0$ decays as the magnetic flux increases and approach zero in
the limit of large magnetic flux (i.e. $\Phi_{c}/\Phi_{0}>0.2$). Due
to the lifting of the degeneracies, the energy spectrum of an
armchair hexagonal dot exhibits an almost linear behavior around
$E=0$ as function of eigenvalue index where, both TBM and continuum
models approximately display the same slope for the linear regime.

\begin{figure}
\centering
\includegraphics[width=8cm]{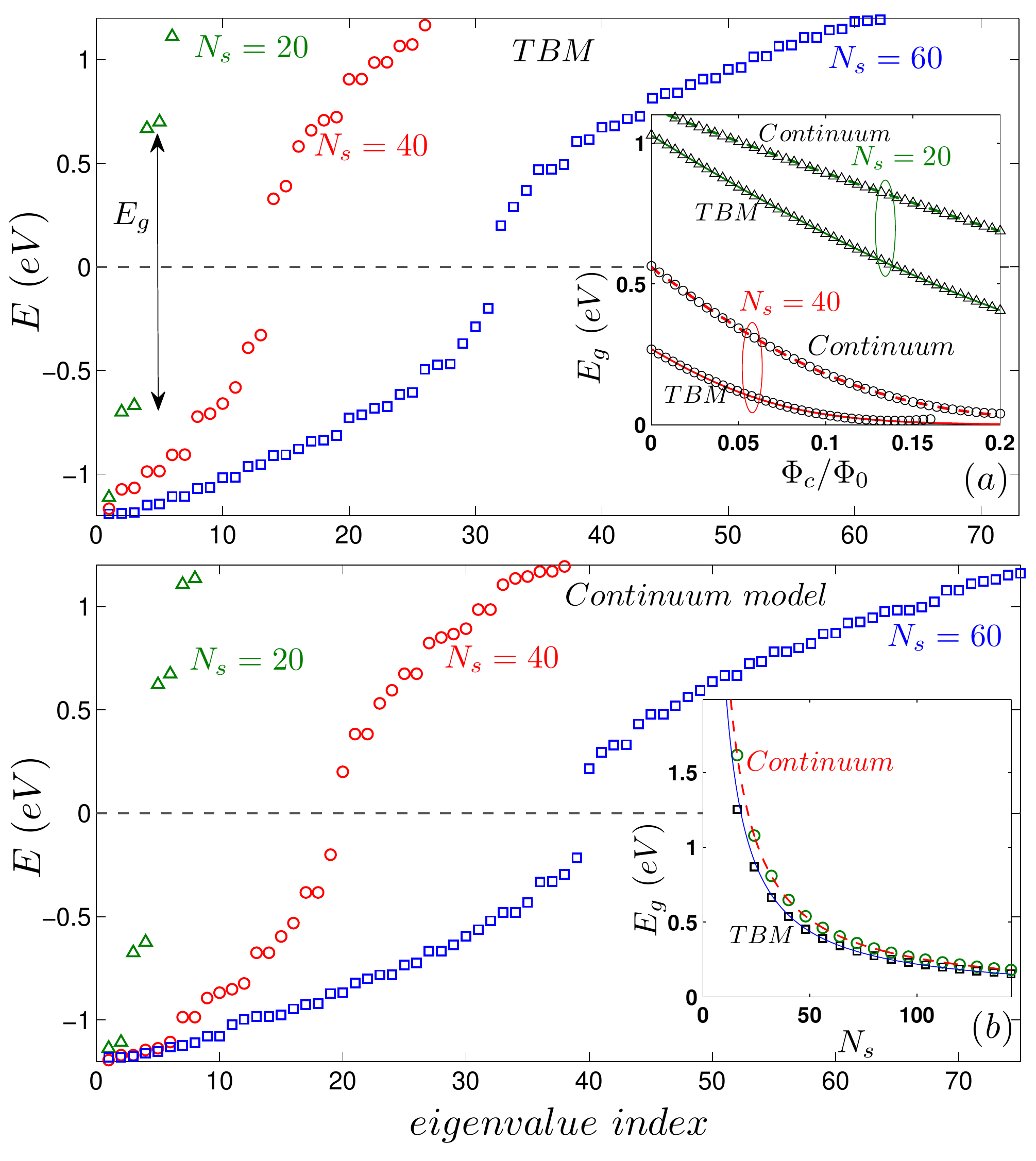}
\caption{The same as Fig. 6 but in the presence of an external
magnetic field $B=50$ T. The inset in panel (a) shows the energy gap
$E_{g}$, obtained by the TBM (solid curves) and continuum model
(dashed curves), as function of the magnetic flux through one carbon
ring $\Phi_{c}$ for $N_{s}=20$ and $N_{s}=40$. The triangle and
circle symbols display Eq. (13) which is fitted to the numerical
results. In the inset of panel (b) $E_{g}$ is shown as function
$N_{s}$ for both TBM (black squares) and continuum models (green
circles) in the presence of an external magnetic field $B=50~T$.}
\label{fig11}
\end{figure}

For triangular graphene flakes under a $B = 50$ T (i.e.
$\Phi_{c}/\Phi_{0}=0.0063$) magnetic field, the energy spectra
obtained by the TBM (a) and the continuum model (b) are shown in
Fig. \ref{fig10}, considering zigzag edges, and Fig. \ref{fig11},
considering armchair edges. As mentioned earlier, the presence of a
magnetic field does not affect the $E = 0$ edge states in the
triangular zigzag flakes, but lifts the degeneracy of the $E \neq 0$
states. The energy gap $E_{g}$ around $E=0$ of triangular flakes is
shown as function of magnetic flux $\Phi_{c}$ in the insets of Fig.
\ref{fig10}(a) and Fig. \ref{fig11}(a) respectively for zigzag (with
$N_{s}=12,24$) and armchair (with $N_{s}=20,40$) edges (circle and
triangle symbols present the fitted results).
$E_{g}^{0,1,2}=(1.12,-1.32,-0.028)$ and
$E_{g}^{0,1,2}=(1.5,-1.77,0.4)$ are the fitting parameters of a
zigzag triangular dot with $N_s=12$ respectively for TBM and
continuum results (see inset of Fig. \ref{fig10}(a)). The fitting
parameters for an armchair dot with $N_{s}=20$ (see inset of Fig.
\ref{fig11}(a)) obtained by TBM and continuum models are
respectively $E_{g}^{0,1,2}=(1.02 ,-3.87,3.83)$ and
$E_{g}^{0,1,2}=(1.12,-2.41,1.2)$. In both zigzag (with $N_s=12$ and
armchair ($N_s=20$) triangular dots the fittings are done for the
range of $0\leq\Phi_{c}/\Phi_{0}\leq 0.2$). In contrast with
hexagonal dots the energy gap of the triangular dots reduces
smoothly (i.e. almost linearly) with increasing the magnetic flux.
Therefore the energy gap is weakly affected by a low magnetic field
in triangular graphene dots. In the inset of Fig. \ref{fig11}(b) the
energy gap is shown as function of $N_{s}$. As in the case of zero
magnetic field $E_{g}$ can be fitted to $E_{g}=\alpha/N_{s}$ as
function of $N_{s}$ (see blue solid and red dashed curves in Fig.
\ref{fig11}(b)). The fitting parameter for $B=50~T$ are
$\alpha\approx 21.87$ eV for TBM and $\alpha\approx 25.9$ eV for the
continuum model which is almost the same as for zero magnetic field
(see Fig. \ref{fig5}), i.e. because of the linear magnetic field
dependence of the energy gap for low magnetic field it does not
affect significantly the dependence of the energy gap on the size of
the armchair triangular graphene dot.

As a matter of fact, tuning the energy gap by adjusting the external
magnetic field is more useful for smaller sizes of the dot, since
the energy gap decays to zero as the size of the dot increases. On
the other hand, due to the small size of the dots considered in
Figs. \ref{fig8} - \ref{fig11}, we need large magnetic field values
(e.g. $B = 50$ T) in order to see its effect on the energy spectrum.
Nevertheless, as the influence of the magnetic field scales with the
magnetic flux through the dot area, similar results will be obtained
for lower magnetic fields in case of a larger graphene dot.

\section{Summary and concluding remarks}
We have presented a theoretical study of triangular and hexagonal
graphene quantum dots, using the two well known models of graphene:
the tight-binding model and the continuum model. For the continuum
model, the Dirac-Weyl equations are solved numerically, considering
armchair, zigzag and infinite mass boundary conditions. A comparison
between the results obtained from the TBM and the Dirac-Weyl
equations show the importance of boundary conditions in finite size
graphene systems, which affects their energy spectra. The results
obtained by the TBM for graphene flakes are only qualitatively
similar to the results from the Dirac-Weyl equation for such systems
considering zigzag and armchair boundary conditions, which shows
that energy values obtained from the continuum model for small
graphene dots may not always be quantitatively reliable.

More specifically, for zigzag hexagonal and triangular dots, the DOS
at $E = 0$ in the absence of a magnetic field is overestimated in
the continuum approach. Similarly, the continuum model also
overestimates the energy gap around $E = 0$ in the armchair case for
both geometries. A good agreement between both models is only
observed for very large dots, as expected, and such agreement is
always better for the triangular case, as compared to the hexagonal
case. The energy spectrum obtained using the continuum model with
infinite mass boundary condition for hexagonal graphene flakes do
not exhibit the same properties as the results obtained with the
armchair or zigzag boundaries (in both TBM and continuum models),
which shows that this type of boundary condition may not give a good
description of finite size hexagonal graphene flakes. On the other
hand, for the triangular case, the results from the continuum model
with infinite mass boundary conditions describe very well the case
of triangular dots with armchair edges.

In the presence of an external magnetic field, the energy levels
obtained by the continuum model with zigzag and armchair boundary
conditions converge to the Landau levels of graphene as the magnetic
field increases, as observed in the TBM. However, many additional
artifact states appear in the continuum model, which do not match
with any TBM result and do not approach any Landau level. Besides,
the influence of an external magnetic field on the gap in the energy
spectra of graphene flakes is particulary different for triangular
and hexagonal dots. The energy gap of the hexagonal flakes (with
$N_{s}\leq 10$) reduces quickly with increasing the magnetic flux,
whereas the gap of the triangular flakes decreases smoothly as the
magnetic flux increases. This feature is observed in both TBM and
continuum model, and suggests that the energy gaps of hexagonal
flakes are more easily controllable by an applied external field, as
compared to the triangular graphene dots.\\

\acknowledgements
This work was supported by the Flemish Science Foundation (FWO-Vl), the Belgian
Science Policy (IAP), the European Science
Foundation (ESF) under the EUROCORES Program
EuroGRAPHENE (project CONGRAN), the Bilateral program between Flanders and Brazil, CAPES
and the Brazilian Council for Research (CNPq).

\end{document}